\documentclass[aps,prb,amsmath,reprint,superscriptaddress]{revtex4-1}
\usepackage{amsmath,txfonts,bm,color,dcolumn,graphicx,physics,multirow}
\begin{document}

\title{Multireference electron correlation methods: Journeys along potential energy surfaces}
\author{Jae Woo Park}
\email{jaewoopark@cbnu.ac.kr}
\affiliation{Department of Chemistry, Chungbuk National University, Chungdae-ro 1, Cheongju 28644, Korea.}
\author{Rachael Al-Saadon}
\affiliation{Department of Chemistry, Northwestern University, 2145 Sheridan Rd., Evanston, IL 60208, USA.}
\author{Matthew K. MacLeod}
\affiliation{Workday, 4900 Pearl Circle East, Suite 100, Boulder, CO 80301, USA.}
\author{Toru Shiozaki}
\affiliation{Department of Chemistry, Northwestern University, 2145 Sheridan Rd., Evanston, IL 60208, USA.}
\affiliation{Quantum Simulation Technologies, Inc., 625 Massachusetts Ave., Cambridge, MA 02139, USA.}
\author{Bess Vlaisavljevich}
\email{Bess.Vlaisavljevich@usd.edu}
\affiliation{Department of Chemistry, University of South Dakota, 414 E. Clark Street, Vermillion, SD 57069, USA.}
\date{\today}

\begin{abstract}
Multireference electron correlation methods describe static and dynamical electron correlation in a balanced way,
and therefore, can yield accurate and predictive results
even when single-reference methods or multiconfigurational self-consistent field (MCSCF) theory fails.
One of their most prominent applications in quantum chemistry is the exploration of potential energy surfaces (PES).
This includes the optimization of molecular geometries, such as equilibrium geometries and conical intersections,
and on-the-fly photodynamics simulations; both depend heavily on the ability of the method to properly explore the PES.
Since such applications require the nuclear gradients and derivative couplings,
the availability of analytical nuclear gradients greatly improves the utility of quantum chemical methods.
This review focuses on the developments and advances made in the past two decades. 
To motivate the readers, we first summarize the notable applications of
multireference electron correlation methods to mainstream chemistry,
including geometry optimizations and on-the-fly dynamics.
Subsequently, we review the analytical nuclear gradient and derivative coupling theories for these methods,
and the software infrastructure that allows one to make use of these quantities in applications.
The future prospects are discussed at the end of this review.
\end{abstract}

\maketitle
\tableofcontents
\section{Introduction}

The need for multiconfigurational methods is motivated by interest in systems where the 
electronic structure cannot be described by a single determinant.
Situations where the single-determinant picture fail are abundant in chemistry. Any chemical reaction involves mixing
the electronic wave function of the reactant and product, and thus a large part of the reaction pathway should be
better described by using more than one electronic configuration.\cite{Shaik2007} Open shell molecules, even at their optimized
ground state equilibrium structures, benefit from multireference approximations to the electronic wave function.\cite{Bally1999,Borden2015,Suaud2014,Trinquier2015}
One example from inorganic chemistry is the chromium--chromium metal--metal bond where
in a simple molecular orbital description
the antibonding $\pi$ and 
$\delta$ orbitals are sufficiently populated in the ground state, reducing the bond order. However, unlike for most other bonds, the metal--metal bond distance is often not a good indicator of
bond strength. The bond distance of the incorrect electronic state may serendipitously be in good agreement with experiment.
By optimizing the geometry for the correct electronic state, we can be predictive.

While the chromium--chromium metal--metal bond is a common example of a ground state where the electronic structure 
is multiconfigurational, excited states require the use of these methods more frequently.
Even in the case that a single determinant is sufficient to describe the electronic structure of the ground state,
it is challenging to describe excited states with multiple electronic excitations with single-reference methods.
In addition, when the ground state intersects with an excited state surface (at a conical intersection),
the isoenergetic states both exhibit multiconfigurational character.
The wave functions in this case are properly described by multistate multireference methods,
since the ground and excited states are treated on an equal footing.
For this reason, photochemical reactions that involve conical intersections should be studied with multireference methods.\cite{Lischka2018CR}

Computing nuclear gradients and derivative couplings is essential for such applications.
Though the nuclear gradients may be obtained by evaluating finite differences between the electronic energies (often referred to as numerical gradients),
it is not practical as it requires many energy evaluations at distorted geometries.
The numerical gradient approach also suffers from rounding errors due to the use of finite precision.
Therefore, the code to calculate analytical nuclear gradients is an important computational tool 
that not only greatly reduces the computational burden but also improves the accuracy.
Historically, nuclear gradient theory for the Hartree--Fock method 
was first formulated as early as the 1960s;\cite{Bratoz1958ColloqIntCNRS,Gerratt1968JCP,Pulay1969MP}
subsequently, nuclear gradient theory for multiconfigurational self-consistent field (MCSCF) theory,
a multiconfigurational extension of Hartree--Fock,
appeared in the late 1970s.\cite{Kato1979CPL,Pulay1977bookchapter} 

For predictive chemical simulations,
one often has to consider dynamical electron correlation.
If the correction for electron correlation is added to the MCSCF reference,
these methods are called multireference (MR) electron correlation theories,
which include multireference configuration interaction (MRCI), coupled cluster (MRCC), and perturbation theory (MRPT).
In fact, analytical nuclear gradient theory for uncontracted MRCI already appeared by the 1980s,\cite{Lengsfield1984JCP,Shepard1987IJQC}
and since that time, has proven to be very useful
(see Secs.~\ref{sec_application_geomopt}~and~\ref{sec_application_dynamics} of this review).
For analytical nuclear gradients for MRCC and MRPT,
there have been two earlier works in the 1990s: one is the two-determinant MRCC gradient in 1994\cite{Szalay1994JCP}
and formulation of the multi-configuration quasi-degenerate second-order perturbation theory (MCQDPT2) gradient in 1998 (though implementation has never been reported).\cite{Nakano1998JCP}
The rest of the significant progress has been made in the last two decades:
for example, the development of an analytical nuclear gradient program for partially internally contracted perturbation theory, PIC-CASPT2,\cite{Celani2003JCP}
quickly became a workhorse for the application of multireference methods in geometry optimizations and dynamics for small molecules.
In 2015, an analytical nuclear gradient program for fully internally contracted perturbation theory, FIC-CASPT2,\cite{MacLeod2015JCP} was developed with the aid of the automatic code generation technique.
Since FIC-CASPT2 is computationally less demanding than PIC-CASPT2,
it has extended the domain of applicability of such simulations; 
most notably, several CASPT2 on-the-fly dynamics studies have been reported in recent literature.\cite{Mai2016JPCL,Park2017JCTC2,Heindl2019CTC,Luzon2019JPCL,Polyak2019JCTC,Gomez2019PCCP}
We will review in the following the recent methodological advances and their applications.

The structure of this review is as follows:
First, we review applications for molecular systems for which nuclear gradients for multireference calculations have played an important role.
In particular, we focus on the geometry optimization
and dynamics simulations using MR methods in Secs.~\ref{sec_application_geomopt}~and~\ref{sec_application_dynamics}, respectively. 
Next, in Sec.~\ref{sec_electronic_structure}, we review multireference electron correlation methods on the basis of which the nuclear gradient programs have been developed.
In Secs.~\ref{sec_gradient_theory}~and~\ref{sec_mr_gradient}, we review the nuclear gradient and derivative coupling theories:
Sec.~\ref{sec_gradient_theory} recapitulates the nuclear gradient theory for single-reference theories and multireference self-consistent field theory;
and Sec.~\ref{sec_mr_gradient} reviews those for multireference electron correlation methods.
We include a thorough review of the Lagrangian (or response function) formalism that makes mathematical derivations straightforward,
which is now the common apparatus of modern gradient theory.
Techniques for making use of nuclear gradients
in geometry optimizations and dynamics are presented in Sec.~\ref{sec_interface}.
Finally, we end with concluding remarks and future prospects in Sec.~\ref{sec_conclusion}.

\section{Applications on Geometry Optimizations}\label{sec_application_geomopt}

\begin{table*}[]
	\caption{List of optimization studies using multireference electron correlation methods highlighted here. The work that includes dynamics simulations using multireference electron correlation methods are excluded and are given in Table~\ref{table:dynamics}.\label{table:optimization}}
	\begin{tabular}{ccccccccc}
	\hline
	Year & System               & ES method\footnotemark[1]      & Basis set              & Active space\footnotemark[2]    & $N_\mathrm{state}$\footnotemark[3] & ES software\footnotemark[4] & Opt. Geom.\footnotemark[5]           & Ref.                           \\\hline
	1984 & $\mathrm{BeH}_2$     & MRCISD         & \footnotemark[6]                      & (4\textit{e},4\textit{o})         & 1      &             & E-T-                 & \onlinecite{Page1984JCP}                    \\
	2001 & $\mathrm{NH}_3$      & MRCISD,MRAQCC  & DZPP, TZPP             & (8\textit{e},7\textit{o}),(8\textit{e},8\textit{o}) & 2      & {{\sc columbus}}\cite{Lischka2001PCCP,columbus1,columbus2}    & E-T-                 & \onlinecite{Lischka2001PCCP}                \\
	2001 & Ethylene             & MRCISD         & cc-pVDZ, cc-pVTZ       & (2\textit{e},2\textit{o})         & 1      & {{\sc columbus}}\cite{Lischka2001PCCP,columbus1,columbus2}    & E---                 & \onlinecite{Lischka2001PCCP}                \\
	2001 & Formaldehyde         & MRCISD,MRAQCC  & aug'-cc-pVXZ\footnotemark[5]           & (6\textit{e},5\textit{o})         & 5,4    & {{\sc columbus}}\cite{Lischka2001PCCP,columbus1,columbus2}    & E---                 & \onlinecite{Dallos2001JCP}                  \\
	2002 & Formaldehyde         & MRCISD         & aug-cc-pVDZ            & CAS(6\textit{e},4\textit{o})+RAS  & 6      & {{\sc columbus}}\cite{Lischka2001PCCP,columbus1,columbus2}    & E---                 & \onlinecite{Lischka2002MP}                  \\
	2003 & \footnotemark[7]                    & MS-CASPT2    & 6-31G*                 & \footnotemark[8]              & \footnotemark[9]    & {{\sc molcas}}\cite{molcasref}      & EC--                 & \onlinecite{Page2003JCC}                    \\
	2003 & Pyrrole & MS-CASPT2 & aug-cc-pVTZ & \footnotemark[8] & \footnotemark[9] & {{\sc molpro}}\cite{MOLPRO} & E--- & \onlinecite{Celani2003JCP} \\
	2004 & Formaldehyde         & MRCISD         & cc-pVDZ,cc-pVTZ        & (6\textit{e},5\textit{o})         & 3      & {{\sc columbus}}\cite{Lischka2001PCCP,columbus1,columbus2}    & -C--                 & \onlinecite{Lischka2004JCP},\onlinecite{Dallos2004JCP}   \\
	2004 & Ethylene             & MRCISD         & cc-pVDZ,cc-pVTZ        & (4\textit{e},4\textit{o})         & 3      & {{\sc columbus}}\cite{Lischka2001PCCP,columbus1,columbus2}    & -C--                 & \onlinecite{Dallos2004JCP}                  \\	
	2004 & Ethylene & MRCISD & aug-cc-pVXZ\footnotemark[5] & (2\textit{e},2\textit{o}) & 3 & {{\sc columbus}}\cite{Lischka2001PCCP,columbus1,columbus2} & EC-- & \onlinecite{Barbatti2004JCP} \\
	2005 & \footnotemark[10]                    & MS-CASPT2      & cc-pVDZ,cc-pVTZ,6-31G* & \footnotemark[8]              & \footnotemark[9]    & {{\sc molcas}}\cite{molcasref}      & EC--                 & \onlinecite{SerranoAndres2005JCP}           \\
	2005 & Ura, Ade             & MRCIS          & cc-pVDZ                & (12\textit{e},9\textit{o})        & 3,5    & {{\sc columbus}}\cite{Lischka2001PCCP,columbus1,columbus2}    & EC-R                 & \onlinecite{Matsika2005JPCA}                \\
	2006 & Butadiene  & MS-CASPT2 & 6-31G** & (4\textit{e},4\textit{o}) & 4 & {{\sc molpro}}\cite{MOLPRO} & EC-- & \onlinecite{Levine2006MP} \\
	2007 & Dioxetane            & MS-CASPT2      & ANO-RCC                & (12\textit{e},10\textit{o})       & 4      & {{\sc molcas}}\cite{molcasref}      & ECTR                 & \onlinecite{DeVico2007JPCA}                 \\
	2008 & \footnotemark[11]                    & MS-CASPT2      & 6-31G**, 6-31G*        & \footnotemark[8]              & \footnotemark[9]    & {{\sc molpro}}\cite{MOLPRO}      & -C--                 & \onlinecite{Levine2008JPCB}                    \\
	2008 & $\mathrm{F}_2\mathrm{+CH}_3\mathrm{SCH}_3$           & CASPT2         & \footnotemark[6]                      & \footnotemark[8]              & 1      & {{\sc molpro}}\cite{MOLPRO}      & E-TR                 & \onlinecite{Lu2008JCP},\onlinecite{Shao2009JCP}          \\
	2009 & Cyclohexadiene       & MS-CASPT2      & 6-31G*                 & (6\textit{e},6\textit{o})         & 3      & {{\sc molpro}}\cite{MOLPRO}      & EC--                 & \onlinecite{Mori2009CPL}                    \\
	2009 & Hydromethoxy radical & SS-CASPT2      & AVTZ, AVQZ             & (11\textit{e},8\textit{o})        & 1      & {{\sc molpro}}\cite{MOLPRO}      & E---                 & \onlinecite{Eisfeld2009JCP}                 \\
	2010 & PSB3                 & MS-CASPT2 & 6-31G*                 & (6\textit{e},6\textit{o})         & 3      & {{\sc molpro}}\cite{MOLPRO}      & EC--                 & \onlinecite{Mori2010JCP}                    \\
	2011 & Benzene              & MS-CASPT2      & 6-31G*                 & (6\textit{e},6\textit{o})         & 3      & \footnotemark[12]      & EC--                 & \onlinecite{Thompson2011FaradayDiscuss}     \\
	2011 & Pyrrole & XMS-CASPT2 & aug-cc-pVTZ & (8\textit{e},7\textit{o}) & 5 & {{\sc molpro}}\cite{MOLPRO} & E--- & \onlinecite{Shiozaki2011JCP3} \\
	2012 & PSB3 & \footnotemark[13] & 6-31G* & (6\textit{e},6\textit{o}) & 2 & \footnotemark[14] & ECT- & \onlinecite{Gozem2012JCTC} \\
	2012 & Ura, Thy, 5F-Ura     & MS-CASPT2      & Sapporo-DZP            & (10\textit{e},8\textit{o})        & 5      & {{\sc molpro}}\cite{MOLPRO}      & ECTR                 & \onlinecite{Yamazaki2012JPCA}               \\
	2013 & Stilbene & XMCQDPT2 &cc-pVTZ & (2\textit{e},2\textit{o}) to (14\textit{e},14\textit{o}) & 2,3 & {{\sc firefly}}\cite{Firefly} & ECTR & \onlinecite{Ioffe2013JCTC} \\
	2013 & Acetophenone & XMCQDPT2 & 6-31+G* & (10\textit{e},9\textit{o}) & 10/6 & {{\sc firefly}}\cite{Firefly} & EC-- & \onlinecite{Huix-Rollant2013PCCP} \\
	2013 & Ethylene             & MS-CASPT2      & 6-31G*                 & (2\textit{e},2\textit{o})         & 3      & {{\sc molpro}}\cite{MOLPRO}      & -C--\footnotemark[15]                & \onlinecite{Mori2013JCTC}                   \\
	2013 & Pyrazine             & XMS-CASPT2     & aug-cc-pVTZ            & (10\textit{e},8\textit{o})        & 2      & {{\sc molpro}}\cite{MOLPRO}      & EC--                 & \onlinecite{Shiozaki2013PCCP}               \\
	2013 & RPSB models          & MS-CASPT2      & ANO-L-VDZP             & (10\textit{e},10\textit{o})       & 1,2    & {{\sc molpro}}\cite{MOLPRO}      & E---                 & \onlinecite{Walczak2013JCTC}                \\
	2014 & \footnotemark[16]                    & MRCISD         & 6-31+G**, 6-31G**      & \footnotemark[8]              & 2      & {{\sc columbus}}\cite{Lischka2001PCCP,columbus1,columbus2}    & -C--                 & \onlinecite{Nikiforov2014JCP}  \\
	2014 & Ethylene cation & XMS-CASPT2 & 6-311G** & (11\textit{e},7\textit{o}) & 4& {{\sc molpro}}\cite{MOLPRO} & ECT- & \onlinecite{Joalland2014JPCL}             \\
	2015 & QOT2 & XMS-CASPT2 & 6-31G* & (10\textit{e},8\textit{o}) & 5 & {{\sc molpro}}\cite{MOLPRO} & E--- & \onlinecite{Chien2015JPCC} \\
	2015 & KFP Chromophore & XMCQDPT2 & cc-pVDZ & (12\textit{e},11\textit{o}) & 2 & {{\sc firefly}}\cite{Firefly} & E--- & \onlinecite{Mironov2015JPCB} \\
	2016 & Thymine & MS-CASPT2 & 6-31G* & (10\textit{e},8\textit{o}) & 2 & {{\sc molcas}}\cite{molcasref} & ECTR &  \onlinecite{SegarraMarti2016Molecules} \\
	2016 & Azaborine            & CASPT2         & 6-31G*                 & (6\textit{e},7\textit{o})         & 1      & {{\sc molpro}}\cite{MOLPRO}      & E---                 & \onlinecite{Edel2016JCC}                    \\
	2016 & \footnotemark[17]                    & XMS-CASPT2     & TZVPP/cc-pVDZ          & \footnotemark[8]              & 2      & {{\sc bagel}}\cite{Shiozaki2018WCMS}       & E---                 & \onlinecite{Vlaisavljevich2016JCTC}         \\
	2017 & Ketoisocytosine      & MRCISD         & 6-31G*                 & (6\textit{e},5\textit{o})         & 4      & {{\sc columbus}}\cite{Lischka2001PCCP,columbus1,columbus2}    & EC--                 & \onlinecite{Hu2017PCCP}                     \\
	2017 & RPSB models          & MS-CASPT2      & cc-pVDZ                & \footnotemark[8]               & 3      & {{\sc molcas}}\cite{molcasref}      & E---                 & \onlinecite{Dokukina2017PhotochemPhotobiol} \\
	2017 & \footnotemark[18]                    & XMS-CASPT2     & cc-pVDZ                & \footnotemark[8]              & \footnotemark[9]    & {{\sc bagel}}\cite{Shiozaki2018WCMS}       & EC--                 & \onlinecite{Park2017JCTC}                   \\
	2017 & \footnotemark[19] & MS-CASPT2 & aug-cc-pVTZ & \footnotemark[8] & \footnotemark[9] & {\sc molcas}\cite{molcasref} & E--- & \onlinecite{Budzak2017JCTC} \\
	2018 & RPSB models\footnotemark[20]          & XMS-CASPT2     & cc-pVXZ\footnotemark[5]                & \footnotemark[8]               & \footnotemark[9]     & {{\sc bagel}}\cite{Shiozaki2018WCMS}       & EC--                 & \onlinecite{Park2018MP}                     \\
	2018 & PSB3                 & XMS-CASPT2     & 6-31G*                 & (6\textit{e},6\textit{o})         & 2      & {{\sc bagel}}\cite{Shiozaki2018WCMS}       & -C--                 & \onlinecite{Sen2018MP}                      \\
	2019 & Ura, Thy cations     & XMS-CASPT2     & ANO-L                  & (13\textit{e},10\textit{o})       & 4      & {{\sc bagel}}\cite{Shiozaki2018WCMS}       & EC--                 & \onlinecite{SegarraMarti2019PCCP}           \\
	2019 & Cyt, 5M-Cyt          & XMS-CASPT2     & ANO-L                  & (13\textit{e},10\textit{o})       & 4      & {{\sc bagel}}\cite{Shiozaki2018WCMS}       & EC--                 & \onlinecite{SegarraMarti2019ChemPhotoChem}  \\
	2019 & 2-SeUra              & MS-CASPT2      & ANO-RCC-VDZP           & (12\textit{e},9\textit{o})        & 4/3    & {{\sc molcas}}\cite{molcasref}      & EC--                 & \onlinecite{Mai2019JCTC1}                   \\
	2019 & \footnotemark[21]                    & XMS-CASPT2     & cc-pVDZ                & \footnotemark[8]              & \footnotemark[9]    & {{\sc bagel}}\cite{Shiozaki2018WCMS}       & EC--                 & \onlinecite{Park2019JCTC2}                  \\
	2019 & \footnotemark[22]                    & MS-CASPT2      & cc-pVDZ                & \footnotemark[8]              & 3      & {{\sc bagel}}\cite{Shiozaki2018WCMS}       & EC--                 & \onlinecite{Park2019JCTC}                   \\ \hline
\end{tabular}
	\footnotetext[1]{Electronic structure method.}
	\footnotetext[2]{For MRCI studies, the active space for generating reference configurations.}
	\footnotetext[3]{Number of states in the SA-CASSCF and the MR correlation method. $a$/$b$ denotes $a$ singlet states and $b$ triplet states. }
	\footnotetext[4]{Electronic structure software, if mentioned.}
	\footnotetext[5]{Type of optimized geometries. E, C, T, and R indicate equilibrium geometries, conical intersection geometries, transition state geometries, and reaction pathways, respectively.}
	\footnotetext[6]{Combined basis functions or various basis functions are used. See original reference.}
	\footnotetext[7]{Butadiene, acrolein, acetone, propenoic acid, diazomethane, pyrrole, PSB3, PSB5.}
	\footnotetext[8]{Active space is chosen for each system. See original reference.}
	\footnotetext[9]{Different number of states used for each system. See original reference.}
	\footnotetext[10]{Formaldehyde, ethylene dimer, PSB3.}
	\footnotetext[11]{Ethylene, formaldehyde, butadiene, stilbene, model GFP chromophore anion (\textit{p}HBI).}
	\footnotetext[12]{Equilibrium geometry using {{\sc molpro}}\cite{MOLPRO} (numerical gradient), and conical intersections using {{\sc molcas}}\cite{molcasref} interfaced with CIOPT.}
	\footnotetext[13]{MS-CASPT2 and XMCQDPT2.}
	\footnotetext[14]{{{\sc molcas}}\cite{molcasref} for MS-CASPT2 and {{\sc firefly}}\cite{Firefly} for XMCQDPT2.}
	\footnotetext[15]{Automated search for conical intersections.}
	\footnotetext[16]{Methyliminium, ethylene, butadiene, PSB3, styrene, ketene, \textit{p}HBI.}
	\footnotetext[17]{Copper corrole, benzophenone.}
	\footnotetext[18]{Ethylene, PSB3, stilbene, model GFP chromophore anion (\textit{p}HBDI).} 
	\footnotetext[19]{Set of small organic compounds. See original reference.}
	\footnotetext[20]{Truncated RPSB chromophores with 3, 4, 5, 6 double bonds.}
	\footnotetext[21]{Iron porphyrin, adenine, \textit{p}HBI.}
	\footnotetext[22]{PSB3, \textit{p}HBI.}
\end{table*}

\subsection{MRCI Applications}

Uncontracted MRCI with singles and doubles (MRCISD) was among the first multireference electron correlation methods to have analytical nuclear gradients implemented.
This led to the first application of the MRCISD analytical nuclear gradient in work by Page \textit{et al.} in 1984,\cite{Page1984JCP}
which was based on the theory by Osamura \textit{et al.} in 1982.\cite{Osamura1982JCP}
In this work, the equilibrium and transition state geometries for the Be + $\mathrm{H}_2$ $\rightarrow$ $\mathrm{BeH}_2$ reaction were optimized using complete active space SCF (CASSCF), MRCISD, and CI with a selected reference.
The barrier for this reaction differs by 4 kcal/mol with and without the CI corrections.

Efficient uncontracted MRCISD analytical nuclear gradients\cite{Shepard1987IJQC,Shepard1992JCP,Lischka2001PCCP,Lischka2002MP} were implemented in {{\sc columbus}}.\cite{Lischka2001PCCP,columbus1,columbus2}
Using {{\sc columbus}}, in 2001 Lischka \textit{et al.}\cite{Lischka2001PCCP} performed geometry optimizations of the S$_1$ equilibrium structure and 
predissociative state of NH$_3$ and the ground state of ethylene with the MRCISD and multireference averaged quadratic coupled cluster (MRAQCC) methods.
The MRAQCC method corrects for size-consistency and can be considered as an intermediate between CI and CC methods.\cite{Lyakh2012CR} The authors were specifically
 interested in the S$_1$ minimum structure of the $^1A^{\prime \prime}_1$ state and the $^1B_1$ dissociative saddle point.
However, since the calculations were performed with 
$C_{2V}$ symmetry, the S$_1$ minimum was computed for the $^1B_1$ state instead. 
The NH bond distances are in very good agreement with experiment,
 while the vibrational frequencies differ by 10 to 40 cm$^{-1}$ between the two methods (MRCISD versus MRAQCC),
although the modes are in reasonable agreement with experiment. The MRAQCC approach was also used to explore the electronic transition in the 
Phillips band for the C$_2$ molecule and the geometry of ethylene was studied with MRCISD. 
The MRCISD and MRAQCC methods were later applied for the excited states of formaldehyde,\cite{Dallos2001JCP} 
where the equilibrium geometries and harmonic frequencies were computed.
Compared to the geometry optimizations based on the numerical gradient with symmetry imposed using MRCID\cite{Hachey1996JMolSpec} and CASPT2,\cite{Merchan1995TCA}
the optimized structure of the $2$ $^1A_1$ state was determined to be a saddle point.
This result also implied the importance of adding dynamical correlation and optimizing in all dimensions.

The applicability of analytical nuclear gradients and derivative couplings with MRCI\cite{Lischka2004JCP,Lischka2002MP} using a state averaged (SA)-CASSCF reference was demonstrated for ethylene and formaldehyde.\cite{Dallos2004JCP}
An extensive study by Barbatti {\it{et al.}} explored the photochemistry of ethylene
by characterizing the excited-state energy surface with MRCISD.\cite{Barbatti2004JCP} The stationary points and conical intersections were computed using 
analytical nuclear gradients and non-adiabatic coupling terms along the crossing seam between the S$_0$ and S$_1$ states, 
which contains all of the conical intersections known for ethylene.\cite{Barbatti2004JCP}
The active space for ethylene was the minimal 2 electrons in 2 orbitals (2{\it{e}},2{\it{o}}) space consisting of the $\pi$ and $\pi^*$ orbitals, although larger reference spaces were tested to ensure that a minimal active space 
was a reasonable choice. Four additional orbitals, the 3s and 3p, were included in other calculations as an auxiliary space to calculate the Rydberg states; however, only single
excitations were allowed from the CAS(2{\it{e}},2{\it{o}}). The optimizations were performed using the analytical MRCI non-adiabatic coupling vectors.\cite{Lischka2004JCP}
The standard approach of optimizing minima and conical intersections with CASSCF and performing single points with MRCI or CASPT2 calculations, previously 
reported by Ben-Nun and Mart\'inez, was in general shown to be quite good;\cite{Ben-Nun2000CP} however, the CASSCF geometries differ most when the dynamical correlation differs significantly 
between two intersecting states. For ethylene, this occurs in the crossing of Rydberg and valence states that is observed during torsion. 
In this case, the CASSCF crossings are approximately 40$^\circ$ larger than those computed with MRCISD/MRCISD+Q. 

Around the same time, Boggio-Pasqua {\it{et al.}} published their strategy for geometry optimizations of the ionic and covalent $\pi\pi^*$ excited state of butadiene and
hexatriene.\cite{Boggio-Pasqua2004JCP} Unlike the other studies in this section,
the focus of the work was to identify best practices for describing the $\pi\pi^*$ excited states in a balanced, accurate way with a lower cost than MRCI or multireference second-order perturbation theory. 
MRCI and CASPT2 geometries were included for comparision. 
They used the restricted active space SCF method (RASSCF) to recover dynamical correlation by including single excitation
from the core $\sigma$ orbital to all of the valence $\pi$ orbitals. This approach was combined with a one-electron basis set for describing valence ionic states. 
The computed geometries were found to be sensitive to the different ways of including dynamical correlation and the authors argued that by using RASSCF, and including all the valence orbitals, the dynamical correlation
was recovered in a more consistent way. RASSCF also had an analytical nuclear gradient which was in its favor.
\begin{figure}[tb]
	\includegraphics[width=0.8\linewidth]{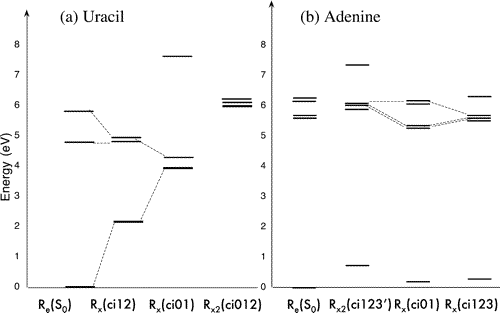}
	\caption{Representation of (a) three and (b) four state calculations at different geometries for two different nucleobases. The relationship between the surfaces at different points, including conical intersections, is shown and was evaluated with MRCI$\sigma \pi$. 
 Reprinted with permission from Ref.~\onlinecite{Matsika2005JPCA}. Copyright 2005 American Chemical Society.
	\label{figure:uracil}}
\end{figure}

A noteworthy series of studies using MRCI methods for chemical systems were conducted for nucleobases by
Barbatti and coworkers\cite{Barbatti2008JACS,Barbatti2010PNAS,Barbatti2011JCP,Barbatti2012JCP}
and Matsika and coworkers.\cite{Matsika2004JPCA,	Matsika2005JPCA,Kistler2007JPCA,Kistler2008JCP,Matsika2011ARPC}
Dynamics simulations were performed in most of the studies by Barbatti and coworkers; 
therefore, we introduce them in Sec.~\ref{sec_application_dynamics}.
An example of optimizations of conical intersections in nucleobases was performed by Matsika {\it{et al.}}
Not only were two-state crossing points, but also three-state crossing points located using MRCIS for the nucleobases uracil and adenine.\cite{Matsika2005JPCA}
In terms of energy, the three-state conical intersection points in adenine were found thermally accessible from the vertical excitation point (Fig.~\ref{figure:uracil}),
and therefore, the role of these conical intersections in radiationless decay was suggested. Three-state conical intersections were also predicted for cytosine and pyrimidinone bases\cite{Kistler2008JCP}, the dynamics of the former were later studied at the CASSCF level of theory.\cite{Richter2012JCPL} 

\begin{figure*}[tb]
	\includegraphics[width=0.6\linewidth]{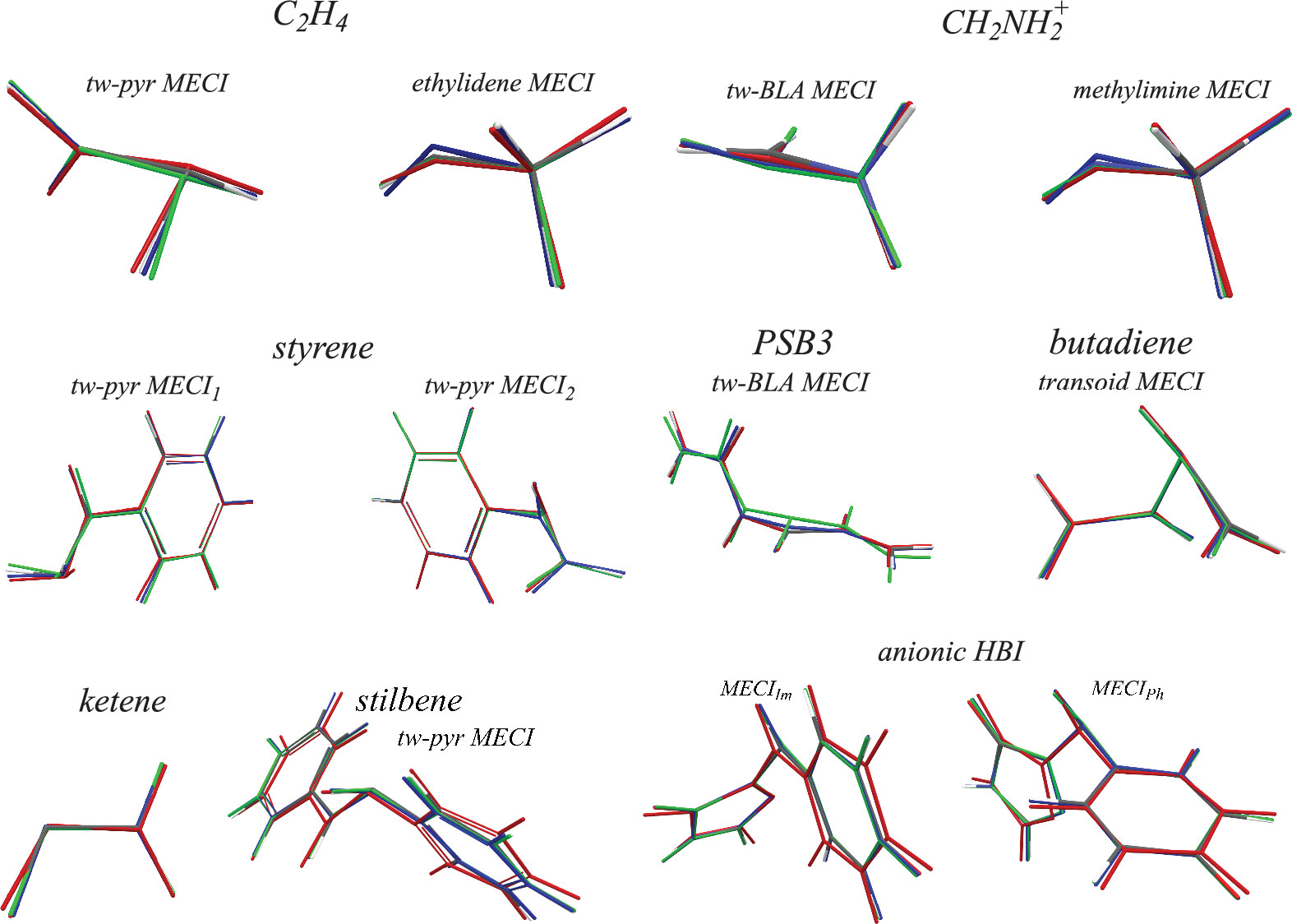}
	\caption{Geometries of organic molecules computed with various methods, including MRCISD, superimposed to demonstrate the accuracy of the optimization.
	The MRCISD results are shown with carbon atoms in grey, hydrogen atoms in white, nitrogen atoms in blue, and oxygen atoms in red.
Reprinted from Ref.~\onlinecite{Nikiforov2014JCP}, with the permission of AIP Publishing.}
\label{figure:mrcisd_ci}
\end{figure*}

A benchmark study for locating the conical intersections of various molecules was performed by Nikiforov \textit{et al.} in 2014,
using MRCISD geometries as the reference.\cite{Nikiforov2014JCP}
They optimized the conical intersections of the molecules shown in Fig.~\ref{figure:mrcisd_ci}.
Semiempirical methods along with density functional theory (DFT) [orthogonalization-corrected model Hamiltonian(OM2/MRCI),\cite{Thiel2014WCMS} spin-flip time-dependent density functional theory (SF-TDDFT),\cite{Shao2003JCP} and state-interaction state-averaged restricted ensemble-referenced Kohn--Sham (SI-SA-REKS)\cite{Filatov2013JCTC}]
were tested against the MRCISD geometries,
and the optimized geometries tended to agree reasonably with the MRCISD references.
Most recently in 2017, the conical intersections of ketoisocytosine were optimized using MRCISD,\cite{Hu2017PCCP}
while the direct dynamics simulations were performed with the cheaper method [ADC(2)].

\subsection{Perturbation Theory Applications: Numerical Gradients}
The majority of the early geometry optimizations using perturbation theories were performed on small molecules using numerical gradients.
One example is the aforementioned CASPT2 optimization of formaldehyde with a symmetry constraint.\cite{Merchan1995TCA}
In 2003, Page and Olivucci 
performed CASSCF and CASPT2 geometry optimizations of small organic molecules in the ground and excited states using {{\sc molcas}} 4.1.\cite{molcasref,Page2003JCC} These calculations used numerical 
gradients to optimize organic chromophores including 1,3-butadiene, acrolein, and two protonated Schiff bases. In some cases, the CASPT2 computed bond distances had large differences from the CASSCF results, approximately 0.01 to 0.02 \AA, 
but the overall trends were generally maintained.
The changes in geometric parameters between double- or triple-$\zeta$ basis sets was significant for both methods. 
If a system was not well-described by CASSCF or strong ionic character was present 
(where presumably the highly localized electron density is also not treated sufficiently well at the CASSCF level), 
the differences tended to be greater. 
Page and Olivucci also optimized the geometry of low-lying conical intersections. CASSCF and CASPT2 yielded fairly large differences in geometry for the {\it{cis}} butadiene S$_0$/
S$_1$ geometry yet this conical intersection is only lower in energy by 4.50 kcal/mol when dynamical correlation is included. 

The first formulation of analytical nuclear gradient theory for uncontracted second-order perturbation theory, the so-called multi-configuration quasi-degenerate second-order PT (MCQDPT2), was presented in 1998,\cite{Nakano1998JCP}
though its implementation and application have not been reported to our knowledge.
Five years later, the analytical nuclear gradient method for partially internally contracted (PIC)-CASPT2 was made publicly available,\cite{Celani2003JCP}
while the fully internally contracted (FIC)-CASPT2 analytical nuclear gradients were not available until they were reported by us in 2015.\cite{MacLeod2015JCP}
Prior to these developments, many of the studies using FIC-CASPT2\cite{Andersson1992JCP} and (X)MCQDPT2\cite{Nakano1993JCP,Granovsky2011JCP}
were performed using numerical gradients.

In 2005, Serrano-Andr\'es {\it{et al.}} used CASPT2 and multistate (MS)-CASPT2 to explore the potential energy surface of LiF, formaldehyde, the ethene dimer, and the penta-2,4-dieniminium cation.\cite{SerranoAndres2005JCP} 
They searched not only for minima, but also for conical intersections along the crossing seams of different exited states. 
The structures were compared with those from CASSCF and MRCI. 
This work discussed practical procedures for using the MS-CASPT2 approach for polyatomic systems.

In 2007, De Vico {\it{et al.}}\cite{DeVico2007JPCA}
used MS-CASPT2 numerical gradients in {{\sc molcas}}\cite{molcasref} to identify the 
the minimum energy pathway for dioxetane dissociation in the ground state.
Since CASSCF and MS-CASPT2 agreed well for the $\mathrm{S}_0$ dissociation pathway,
dissociation along the $\mathrm{S}_1$ and $\mathrm{T}_1$ states
was calculated using an MS-CASPT2//CASSCF protocol
where the geometries were optimized with CASSCF,
but the energy was calculated with MS-CASPT2.
This was also the case for the smallest model retinal protonated Schiff base (RPSB), the penta-2,4-dieniminium cation (PSB3).
PSB3 was also thoroughly studied in work by Gozem {\it{et al.}},\cite{Gozem2012JCTC}
where the structure of PSB3 was optimized with MS-CASPT2 and XMCQDPT2 using numerical gradients.
They also observed that using a MRPT2//CASSCF protocol yielded similar quality results to optimizing with MRPT2.

The photoisomerization of stilbene has been extensively studied with XMCQDPT2.\cite{Ioffe2013JCTC,Doryakov2012JCP}
The equilibrium geometries were optimized and their vibrational frequencies were computed. Additionally, the reaction pathways between these structures were carefully studied.
The CASSCF and XMCQDPT2 vibrational frequencies and the location of the stationary points in the excited states differed significantly.
For example,
the CASSCF surface was found to have an energy minimum on the $\mathrm{S}_1$ surface near the \textit{cis}-stilbene geometry, 
while the XMCQDPT2 calculation could not locate this point.
A critical difference between CASSCF and XMCQDPT2 results was also found for 2-acetyl-4-(p-hydroxybenzyledene)-1-methyl-5-imidazolone (AHBMI),
the kindling fluorescent protein (KFP) chromophore.\cite{Mironov2015JPCB}
The difference in the equilibrium structures obtained by CASSCF and XMCQDPT2 was about 0.05~\AA,
and the emission energies differed by 0.12 eV (difference between XMCQDPT2//XMCQDPT2 and XMCQDPT2//CASSCF) or 0.85 eV (difference between XMCQDPT2//XMCQDPT2 and CASSCF//CASSCF).

Unlike most of the works that report CASPT2 geometry optimization that were focused on organic molecules,
a series of papers by Andrews and coworkers
\cite{Andrews2008ACIE,Cho2009Organometallics1,Cho2009Organometallics2,Wang2010ChemComm,Wang2011InorgChem,Andrews2011EurJInorgChem,Wei2019InorgChem}
explored the electronic structure of small actinide and transition metal complexes in noble gas matrices.
These studies demonstrated the importance and applicability of CASPT2 for geometry optimization and harmonic vibrational frequencies calculations,
using numerical gradients and numerical Hessian calculations available in {{\sc molcas}}.\cite{molcasref} 
The species that were studied include N$\equiv$U$\mathrm{F}_3$ and P$\equiv$U$\mathrm{F}_3$;\cite{Andrews2008ACIE} FB=Th$\mathrm{F}_2$;
\cite{Wang2010ChemComm} N$\equiv$U=N--H;\cite{Wang2011InorgChem} SUO and $\mathrm{US}_2$;\cite{Andrews2011EurJInorgChem}
halomethanes with laser-ablated nickel, palladium, and platinum; \cite{Cho2009Organometallics1,Cho2009Organometallics2} and 
3d transition metal oxyfluoride molecules.\cite{Wei2019InorgChem} 

The numerical gradient implementations of CASPT2 and XMCQDPT2 remain in widespread use for optimizing molecular structures.
Notable examples include locating equilibrium geometries and conical intersections with MS-CASPT2 in 2011,\cite{Thompson2011FaradayDiscuss}
locating equilibrium geometries and conical intersection in acetophenone with XMCQDPT2 in 2013,\cite{Huix-Rollant2013PCCP}
locating equilibrium geometries, conical intersections, transition state geometries, and reaction pathways of thymine with MS-CASPT2 in 2016,\cite{SegarraMarti2016Molecules}
locating equilibrium geometries of various RPSB models in 2017,\cite{Dokukina2017PhotochemPhotobiol}
generating a benchmark set of excited-state geometries of organic molecules with MS-CASPT2 in 2017,\cite{Budzak2017JCTC}
and optimizing equilibrium structures and crossing points in 2-selenouracil in 2019.\cite{Mai2019JCTC1}

\subsection{Perturbation Theory Applications: Analytical Nuclear Gradients}

\begin{figure}[tb]
	\includegraphics[width=0.8\linewidth]{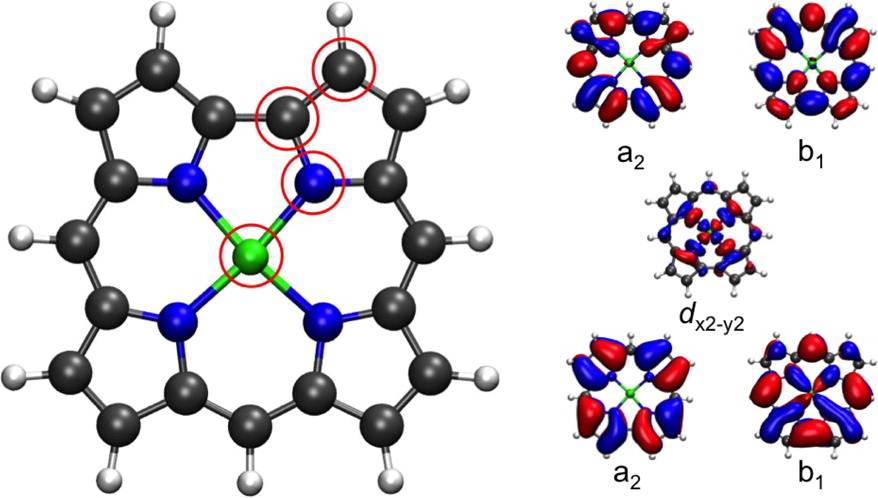}
	\caption{Geometry and active orbitals of copper corrole complex optimized with FIC-CASPT2 analytical gradient.
Reprinted with permission from Ref.~\onlinecite{Vlaisavljevich2016JCTC}. Copyright 2016 American Chemical Society.
		\label{figure:coppercorrole}}
\end{figure}

In this section, we will highlight some the studies that reported geometry optimization utilizing the
PIC-CASPT2 analytical nuclear gradient program in {\sc molpro}\cite{Celani2003JCP,Mori2009CPL,Shiozaki2011JCP3,Gyorffy2013JCP,MOLPRO} and
the FIC-CASPT2 counterpart recently developed in {{\sc bagel}}.\cite{bagel,MacLeod2015JCP,Vlaisavljevich2016JCTC,Park2017JCTC,Park2017JCTC2}
In the original work by Celani and Werner that reported the PIC-CASPT2 nuclear gradient algorithm,
they studied the excited state geometries of pyrrole.\cite{Celani2003JCP}
In this study, the equilibrium geometries of the excited valence states were found to be non-planar,
in stark contrast to the earlier MS-CASPT2 optimization which imposed a planar constraint.\cite{Roos2002JCP}
Pyrrole was also used as an example to demonstrate geometry optimization with XMS-CASPT2.\cite{Shiozaki2011JCP3}
Single-state CASPT2 optimizations (including ground and excited states) have been shown to perform well for
bond formation/ breaking or energy transfer problems for a variety of chemical systems.\cite{Lu2008JCP,Shao2009JCP,Chien2015JPCC}
Furthermore, CASPT2 geometries are often used to benchmark other computational methods,
including those generated under extreme conditions or not observed in experiment.\cite{Edel2016JCC,Eisfeld2009JCP}
These studies include optimizations along the reaction pathway for the $\mathrm{F}_2 + \mathrm{CH}_3 \mathrm{SCH}_3$ reaction;\cite{Lu2008JCP,Shao2009JCP}
ground-state optimization of the hydromethoxy radical;\cite{Eisfeld2009JCP}
geometries and excitation energies for RPSB models for benchmarking the performance of CASSCF, CC2, MP2, and DFT;\cite{Walczak2013JCTC}
the potential energy landscape search of singlet fission material candidate tetracyanoquinodimethane bithiophene (QOT2);\cite{Chien2015JPCC}
and the equilibrium geometry, frequency calculation, and reaction pathway for the ring-opening reaction of azaborine.\cite{Edel2016JCC}
The analytical nuclear gradient for FIC-CASPT2 has been used for the copper corrole complex (Fig.~\ref{figure:coppercorrole}),\cite{Vlaisavljevich2016JCTC}
where a saddled geometry was obtained for the S$_0$ state while the S$_1$ state is planar.

\begin{figure}[tb]
	\includegraphics[width=1.0\linewidth]{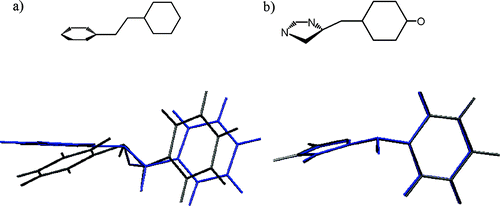}
	\caption{Optimized structures for the $\mathrm{S}_1$/$\mathrm{S}_0$ MECI of (a) stilbene and (b) GFP. Structures from CAS (gray) 
and MS-CASPT2 (blue) are superimposed. Reprinted with permission from Ref.~\onlinecite{Levine2008JPCB}. Copyright 2008 American Chemical Society.
		\label{figure:martinezmeci}}
\end{figure}

In addition to optimizing equilibrium geometries, CASPT2 analytical nuclear gradient theory can be used to locate conical intersections.
In the search for conical intersections using CASPT2, 
work by Levine \textit{et al.} in 2006 and 2008 is notable,\cite{Levine2006MP,Levine2008JPCB}
because these authors introduced a methodology to locate the conical intersection 
[minimal energy conical intersection (MECI) and minimal distance conical intersection (MDCI)] without a derivative coupling vector 
since the analytical interstate coupling vector was not implemented for CASPT2 until 2009\cite{Mori2009CPL}
(note, however, that a scheme for estimating CASPT2 nonadiabatic coupling from CASSCF wave functions was presented by Tao {\it et al.} in the context of dynamics\cite{Tao2009JPCA}).
The conical intersections of ethylene, formaldehyde, butadiene, stilbene, and model GFP chromophore anion (\textit{p}HBI) were optimized.
The CASSCF and CASPT2 optimized MECI geometries exhibited considerable differences for stilbene,
particularly in the degree of pyramidalization (Fig.~\ref{figure:martinezmeci}). 

After the implementation of the interstate coupling vector for CASPT2,\cite{Mori2009CPL}
standard methods for locating conical intersections (which are described in Sec.~\ref{sec_interface}),
such as the gradient projection method,\cite{Bearpark1994CPL}
were applied in conjunction with CASPT2.
The MECI was located for the ring-opening reaction of cyclohexadiene,\cite{Mori2009CPL}
in which the largest deviation in bond lengths between SA-CASSCF and MS-CASPT2 conical intersections was about 0.07~\AA.
Not only were the geometries different, but the overlap ($\mathbf{g}\cdot\mathbf{h}/|\mathbf{g}||\mathbf{h}|$) between the gradient difference vector $\mathbf{g}$ and interstate coupling vector $\mathbf{h}$, 
which are two vectors that characterize the conical intersections, were also inconsistent, with values of 0.335 (SA-CASSCF) and 0.941 (MS-CASPT2).
This implies that the PES topology around the conical intersection changed considerably with the inclusion of dynamical correlation.
Furthermore, conical intersections were optimized on the free energy surface of PSB3\cite{Mori2010JCP}
by combining the free energy within the linear response free energy (LRFE) framework\cite{Sato1996JCP,Ten-no1994JCP,Yamazaki2005JCP} with the CASPT2 gradient and interstate coupling.
When compared to the CASSCF result,
the conical intersections associated with the twisting of the C=N bond became energetically disfavored,
and the C=N bond length in the CASSCF and CASPT2 equilibrium geometries differed by 0.06~\AA.
Conical intersection searches were also performed using PIC-CASPT2,
and the first applications of automated conical intersection search for ethylene\cite{Mori2013JCTC} should also be highlighted,
as this work demonstrates the use of multireference electron correlation methods for exploring conical intersection seams.

The correct topology of the PES near the conical intersection can be obtained using XMCQDPT2 and XMS-CASPT2, which are invariant with respect to rotations among the states as proposed by Granovsky.\cite{Granovsky2011JCP,Shiozaki2011JCP3}
After the development of XMS-CASPT2 nuclear gradients with partial\cite{Shiozaki2011JCP3} and full\cite{Vlaisavljevich2016JCTC} internal contraction,
several conical intersection optimizations have been performed using this method.
One of the earliest examples is the photodynamics simulations of the ethylene cation by Joalland \textit{et al.},\cite{Joalland2014JPCL}
in which three-state MECIs are optimized as well as two-state MECIs.
The RPSB model chromophores have been extensively tested to understand the effect of chromophore truncation,\cite{Park2018MP}
and it turns out that XMS-CASPT2 favors one of two available isomerization channels.
Additionally, the conical intersections of nucleic acids, 
the uracil and thymine cations\cite{SegarraMarti2019PCCP} as well as cytosine,\cite{SegarraMarti2019ChemPhotoChem} have also been optimized using XMS-CASPT2.
The three-state conical intersections in the uracil cation are separated by including the PT2 correction,
with significant changes in the $\mathrm{D}_2$/$\mathrm{D}_1$ conical intersection (up to 0.10~\AA~of change in the bond lengths).\cite{SegarraMarti2019PCCP}

\section{Applications on Dynamics}\label{sec_application_dynamics}

\begin{table*}[]
	\caption{List of the dynamics studies using the multireference electron correlation methods highlighted here.\label{table:dynamics}}
	\begin{tabular}{ccccccccccc}
		\hline
Year & System           & ES method & Basis set    & Active space\footnotemark[1] & $N_\mathrm{state}$\footnotemark[2]          & ES software\footnotemark[3] & $N_\mathrm{traj}$\footnotemark[4] & NAD method\footnotemark[5]  & NAD software\footnotemark[6]  & Ref.                                         \\\hline
2006 & Silaethylene & MRCISD & 6-31G* & (2\textit{e},2\textit{o}) & 3 & {{\sc columbus}}\cite{Lischka2001PCCP,columbus1,columbus2} & 100 & FSSH & & \onlinecite{Zechmann2006CPL} \\
2007 & Methyl salicylate & SS-CASPT2 & 6-31G* & (2\textit{e},2\textit{o}) & 2 & {{\sc molpro}}\cite{MOLPRO} & 5 & AIMS & {{\sc molpro}}\cite{MOLPRO} & \onlinecite{Coe2007JPCA} \\
2007 & \footnotemark[12] & \footnotemark[13] & 6-31G* & \footnotemark[14] & \footnotemark[15] & {{\sc columbus}}\cite{Lischka2001PCCP,columbus1,columbus2} & \footnotemark[16] & FSSH & {{\sc Newton-X}}\cite{Barbatti2014WIREs,newtonx} & \onlinecite{Barbatti2007JPhotochemPhotobiolA} \\
2008 & Adenine & MRCIS & 6-31G*/3-21G & (6\textit{e},4\textit{o}) & 4 & {{\sc columbus}}\cite{Lischka2001PCCP,columbus1,columbus2} & 60 & DC-FSSH\footnotemark[7]& {{\sc Newton-X}}\cite{Barbatti2014WIREs,newtonx} & \onlinecite{Barbatti2008JACS} \\
2009 & Pyrrole          & MRCISD                      & aug'-cc-pVDZ & (4\textit{e},5\textit{o})      & 5                     & {{\sc columbus}}\cite{Lischka2001PCCP,columbus1,columbus2}                      & 90                 & DC-FSSH                      & {{\sc Newton-X}}\cite{Barbatti2014WIREs,newtonx}                       & \onlinecite{Vazdar2009MP}   \\
2009 & Ethylene & MS-CASPT2 & 6-31G* & (2\textit{e},2\textit{o}) & 3& {{\sc molpro}}\cite{MOLPRO} & 13 & AIMS & {{\sc molpro}}\cite{MOLPRO} & \onlinecite{Tao2009JPCA} \\
2010 & Ade, Gua & MRCIS & 6-31G* & \footnotemark[14] & \footnotemark[15] & {{\sc columbus}}\cite{Lischka2001PCCP,columbus1,columbus2} & 60 & DC-FSSH & {{\sc Newton-X}}\cite{Barbatti2014WIREs,newtonx} & \onlinecite{Barbatti2010PNAS} \\
2011 & Ethylene & MS-CASPT2 & 6-31G* & (2\textit{e},2\textit{o}) & 3& {{\sc molpro}}\cite{MOLPRO} & 44 & AIMS & {{\sc molpro}}\cite{MOLPRO} & \onlinecite{Tao2011JCP},\onlinecite{Allison2012JCP} \\
2011 & Guanine & MRCIS & 6-31G* & (10\textit{e},7\textit{o}) & 3 & {{\sc columbus}}\cite{Lischka2001PCCP,columbus1,columbus2} & 60 & DC-FSSH & {{\sc Newton-X}}\cite{Barbatti2014WIREs,newtonx} & \onlinecite{Barbatti2011JCP} \\
2012 & Ethylene & MS-CASPT2 & 6-31G* & (2\textit{e},3\textit{o}) & 5 & {{\sc molpro}}\cite{MOLPRO} & 37 & AIMS & {{\sc molpro}}\cite{MOLPRO} & \onlinecite{Mori2012JPCA} \\
2012 & CPD, $\mathrm{Me}_4$-CPD & MS-CASPT2 & 6-31G** & (4\textit{e},4\textit{o}) & 3 & {{\sc molpro}}\cite{MOLPRO} & 40 each & AIMS & {{\sc molpro}}\cite{MOLPRO} & \onlinecite{Kuhlman2012FaradayDiscuss} \\
2012 & Adenine & MRCIS & 6-31G* & (6\textit{e},4\textit{o}) & 5 & {{\sc columbus}}\cite{Lischka2001PCCP,columbus1,columbus2} & 60 & DC-FSSH & {{\sc Newton-X}}\cite{Barbatti2014WIREs,newtonx} & \onlinecite{Barbatti2012JCP} \\
2012 & Gua in DNA & MRCIS & 6-31G* & (10\textit{e},7\textit{o}) & 4& {{\sc columbus}}\cite{Lischka2001PCCP,columbus1,columbus2} & 53 & DC-FSSH & {{\sc Newton-X}}\cite{Barbatti2014WIREs,newtonx} & \onlinecite{Zeleny2012JACS} \\
2013 & PSB3, MePSB3 & MRCIS & 6-31G & (4\textit{e},5\textit{o}) & 3 & {{\sc columbus}}\cite{Lischka2001PCCP,columbus1,columbus2} & 200 & DC-FSSH & {{\sc Newton-X}}\cite{Barbatti2014WIREs,newtonx} & \onlinecite{Ruckenbauer2013JPCA} \\
2014 & $\mathrm{SO}_2$ & MRCIS & ANO-RCC-VDZP & (6\textit{e},6\textit{o}) & 4/3 & {{\sc columbus}}\cite{Lischka2001PCCP,columbus1,columbus2} & 44/111\footnotemark[10] & SHARC-FSSH & {\sc sharc}\cite{Mai2018WCMS,sharc} & \onlinecite{Mai2014JCP} \\
2014 & $\mathrm{Me}_6$-CPD & MS-CASPT2 & 6-31G** & (4\textit{e},4\textit{o}) & 3 & {{\sc molpro}}\cite{MOLPRO} & 38 & AIMS & {{\sc molpro}}\cite{MOLPRO} & \onlinecite{Wolf2014PCCP} \\
2015 & Cyclohexadiene   & MS-CASPT2                   & cc-pVDZ      & (8\textit{e},8\textit{o})      & 3                     & {{\sc molpro}}\cite{MOLPRO}                        & 87                 & ZNSH         &                                & \onlinecite{Ohta2015CP}     \\
2016 & Sulfine & MS-CASPT2 & 6-31G* & (4\textit{e},3\textit{o}) & 2 & {{\sc molpro}}\cite{MOLPRO} & 102 & AIMS & {{\sc molpro}}\cite{MOLPRO} & \onlinecite{Mignolet2016ACIE} \\
2016 & PSB3 & MS-CASPT2 & 6-31G & (6\textit{e},6\textit{o}) & 3 & {{\sc molpro}}\cite{MOLPRO} & 46 & AIMS & {{\sc molpro}}\cite{MOLPRO} & \onlinecite{Liu2016JPCB} \\
{2016} & 2-thiocytosine & MRCIS & cc-pVDZ & (6\textit{e},5\textit{o}) & 4/2 & {{\sc columbus}}\cite{Lischka2001PCCP,columbus1,columbus2} & 137 & SHARC-FSSH & {{\sc sharc}}\cite{Mai2018WCMS,sharc} & \onlinecite{Mai2016NatCommun} \\
2016 & 2-thiouracil       & MS-CASPT2                   & cc-pVDZ      & (12\textit{e},9\textit{o})     & 4/3 & {{\sc molcas}}\cite{molcasref}                        & 92(44)\footnotemark[8]           & SHARC-FSSH                   & {{\sc sharc}}\cite{Mai2018WCMS,sharc}                          & \onlinecite{Mai2016JPCL}    \\
2017 & \footnotemark[9]    & XMS-CASPT2                  & cc-pVDZ      & \footnotemark[14]  & 3                     & {{\sc bagel}}\cite{Shiozaki2018WCMS}      & \footnotemark[16] & DC-FSSH                      & {{\sc Newton-X}}\cite{Barbatti2014WIREs,newtonx}                       & \onlinecite{Park2017JCTC2}  \\
2017 & 1,2,3-thiadiazole & MS-CASPT2 & 6-31G* & (10\textit{e},8\textit{o}) & 3 & {{\sc molpro}}\cite{MOLPRO} & 150(144) & ZNSH & {\sc gtsh} & \onlinecite{Liu2017JCP} \\
2017 & [2+2] ethylene & SS-CASPT2 & 6-31G* & (4\textit{e},4\textit{o}) & 1 & {{\sc molpro}}\cite{MOLPRO} & 200 & BOMD & {\sc abin} & \onlinecite{Muchova2017IJQC} \\
{2017} & 5-azacytosine & MRCIS & def2-SVP & (6\textit{e},5\textit{o}) & 3/2 & {{\sc columbus}}\cite{Lischka2001PCCP,columbus1,columbus2} & 146 & SHARC-FSSH & {\sc sharc}\cite{Mai2018WCMS,sharc} & \onlinecite{Borin2017PCCP} \\
{2017} & 5-bromouracil & MRCIS & cc-pVDZ-DK & (10\textit{e},8\textit{o}) & 4/3 & {{\sc columbus}}\cite{Lischka2001PCCP,columbus1,columbus2} & 93 & SHARC-FSSH & {\sc sharc}\cite{Mai2018WCMS,sharc} & \onlinecite{Peccati2017PhilTrans} \\
2018 & \textit{trans-}1,3-butadiene & MS-CASPT2 & 6-31G** & (4\textit{e},4\textit{o}) & 3 & {{\sc molpro}}\cite{MOLPRO} & 25 & AIMS & {{\sc molpro}}\cite{MOLPRO} & \onlinecite{Glover2018JCP} \\
2019 & Pyrrole & XMS-CASPT2 & ANO-L & (8\textit{e},8\textit{o}) & 7 & {{\sc bagel}}\cite{Shiozaki2018WCMS} & 23 & DC-FSSH & {{\sc sharc}}\cite{Mai2018WCMS,sharc} & \onlinecite{Heindl2019CTC} \\
2019 & $\mathrm{CH}_3 \mathrm{OH}^{2+}$ & XMS-CASPT2 & aug-cc-pVDZ & (8\textit{e},8\textit{o}) & 7 & {{\sc bagel}}\cite{Shiozaki2018WCMS} & 700 & DC-FSSH & {{\sc Newton-X}}\cite{Barbatti2014WIREs,newtonx} & \onlinecite{Luzon2019JPCL} \\
2019 & Thioformaldehyde & MS-CASPT2                   & ANO-RCC-VDZP & (10\textit{e},6\textit{o})     & 2/2 & {{\sc molcas}}\cite{molcasref}                        & 200                & SHARC-FSSH                   & {{\sc sharc}}\cite{Mai2018WCMS,sharc}                          & \onlinecite{Mai2019JCTC2}   \\
2019 & Cyclohexadiene   & XMS-CASPT2                  & cc-pVDZ      & (6\textit{e},6\textit{o})      & 3                     & {{\sc bagel}}\cite{Shiozaki2018WCMS}                         & 136                & DC-FSSH                      & {{\sc Newton-X}}\cite{Barbatti2014WIREs,newtonx}                       & \onlinecite{Polyak2019JCTC} \\
2019 & 1,1-difluoroethylene & MS-CASPT2 & aug-cc-pVDZ & (2\textit{e},2\textit{o})/(2\textit{e},6\textit{o})\footnotemark[14] & 3/11\footnotemark[14] & {{\sc molcas}}\cite{molcasref} & {152/121}\footnotemark[11] & SHARC-FSSH & {{\sc sharc}}\cite{Mai2018WCMS,sharc}     & \onlinecite{Gomez2019PCCP} \\
\hline
	\end{tabular}
\footnotetext[1]{For MRCI studies, the active space for generating reference configurations.}
\footnotetext[2]{Number of states in SA-CASSCF and MR correlation method. $a$/$b$ denotes $a$ singlet states and $b$ triplet states, {except for ref.~\onlinecite{Gomez2019PCCP}}. }
\footnotetext[3]{Electronic structure software.}
\footnotetext[4]{Number of dynamics trajectories. For AIMS trajectories, the number of initial conditions.}
\footnotetext[5]{Nonadiabatic dynamics method.}
\footnotetext[6]{Nonadiabatic dynamics software, if mentioned.}
\footnotetext[7]{decoherence-corrected fewest-switches surface hopping.}
\footnotetext[8]{The number in parentheses indicates the number of trajectories that did not fail.}
\footnotetext[9]{\textit{p}HBI, the GFP model chromophore, 4-(p-hydroxybenzylidene)-5-imidazolinone anion and adenine in water.}
\footnotetext[10]{44 in singlet manifold and 111 in singlet-triplet manifold.}
\footnotetext[11]{Two simulation conditions were used in this work.}
\footnotetext[12]{Methaniminium and PSB3.}
\footnotetext[13]{MRCISD and MRCIS.}
\footnotetext[14]{Active space is chosen for each system. See original reference.}
\footnotetext[15]{Different number of states used for each system. See original reference.}
\footnotetext[16]{Different number of trajectories used for each system. See original reference.}
\end{table*}

Complex photochemical processes can be simulated by coupling multireference gradient theory to nonadiabatic dynamics methods.
Recent reviews on mixed quantum-classical\cite{Curchod2018CR, CrespoOtero2018CR} nonadiabatic dynamics include details about dynamics simulations with analytical gradients.
Significant effort in software development has made it possible to interface the electronic structure and dynamics software packages required to perform these simulations.
We highlight the results obtained by dynamics simulations with multireference nuclear gradient theory and emphasize the dynamics software employed.
A list of applications involving dynamics reviewed is given in Table~\ref{table:dynamics} along with other relevant details such as the software package, active space, dynamics method, and so on.

\subsection{Uncontracted MRCI Applications}

As was the case for geometry optimizations, the first examples of nonadiabatic dynamics using a multireference correlation method employed uncontracted MRCI,
due to the availability of the analytical gradient in the {{\sc columbus}}\cite{Lischka2001PCCP,columbus1,columbus2} program.
In 2006, Zechmann {\it{et al.}}\cite{Zechmann2006CPL} used MRCISD and the fewest-switch surface-hopping (FSSH) algorithm
to study the dynamics of silaethylene, which was chosen as a model polar $\pi$ bond,
and unconvered the multiple relaxation pathways from the S$_1$ excited state. 

Beginning in 2007, the {{\sc Newton-X}}\cite{Barbatti2014WIREs,newtonx} program has provided a modular approach by separating the electronic structure and dynamics programs.\cite{Barbatti2007JPhotochemPhotobiolA} 
While {{\sc columbus}},\cite{Lischka2001PCCP,columbus1,columbus2} has been primarily used in conjunction with {{\sc Newton-X}}, {\sc turbomole} and {\sc mopac} have been interfaced from the first version of the program.
This program was used to study the methaniminium cation with MRCISD and the retinal model PSB3 cation with MRCIS in {{\sc columbus}}.\cite{Lischka2001PCCP,columbus1,columbus2}

\begin{figure}[tb]
	\includegraphics[width=0.8\linewidth]{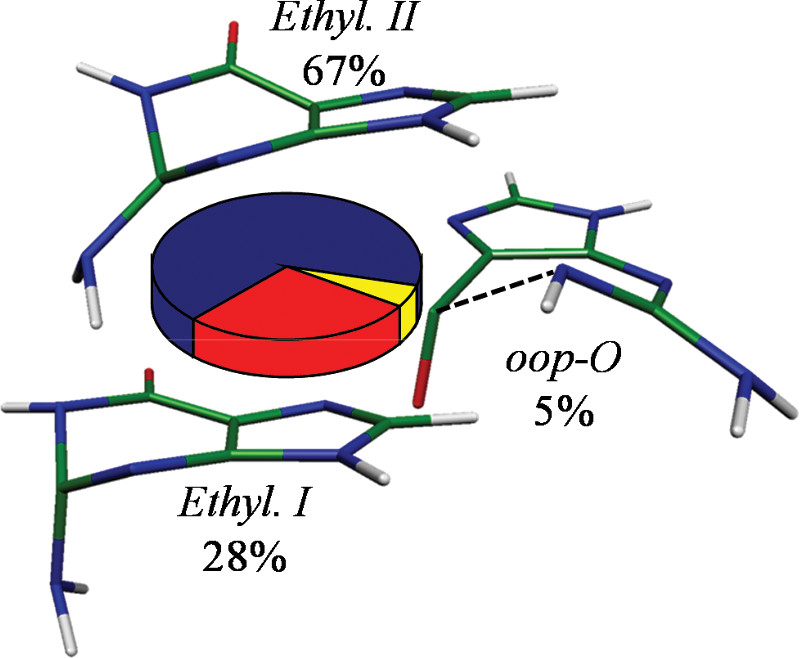}
	\caption{Optimized structures for $\mathrm{S}_1$/$\mathrm{S}_0$ conical intersections of guanine with the percentages of the relative populations from MRCIS dynamics in Ref.~\onlinecite{Barbatti2011JCP}. Reprinted from Ref.~\onlinecite{Barbatti2011JCP}, with the permission of AIP Publishing.	\label{figure:barbattiguanine}}
\end{figure}

In the previous section (Ref.~\ref{sec_application_geomopt}), 
we presented MRCI studies of nucleobases, and many of these studies also include dynamics simulations 
using FSSH.\cite{Barbatti2008JACS,Barbatti2010PNAS,Barbatti2011JCP,Barbatti2012JCP,Zeleny2012JACS}
The first excited-state dynamics study of adenine was reported in 2008 by Barbatti and Lischka.\cite{Barbatti2008JACS}
They investigated nonradiative decay within the singlet manifold using MRCIS with {{\sc columbus}}\cite{Lischka2001PCCP,columbus1,columbus2} 
coupled to the {{\sc Newton-X}}\cite{Barbatti2014WIREs,newtonx} surface-hopping dynamics code.
These simulations included the decoherence effect using the formalism developed by Granucci and Persico.\cite{Granucci2007JCP} 
The study showed two distinct steps including a short relaxation time from S$_3$ to S$_2$ to S$_1$ and a longer relaxation time from S$_1$ to S$_0$. 
In another comprehensive study of the nucleobases (adenine, guanine, cytosine, thymine, and uracil),\cite{Barbatti2010PNAS}
the simulations of adenine and guanine were performed with the MRCIS method,
while the others were simulated with CASSCF.
The deactivation pathway that they found for adenine and guanine is a barrierless path to the conical intersection seam.
Subsequent simulations with MRCIS for adenine\cite{Barbatti2012JCP} and guanine\cite{Barbatti2011JCP}
have shown
that the C2-puckering of these nucleobases (Fig.~\ref{figure:barbattiguanine}) is the main decay pathway. 
In 2012, Zelen{\'y} \textit{et al.} simulated guanine deactivation in the DNA environment
using mixed quantum mechanics--molecular mechanics (QM/MM),\cite{Zeleny2012JACS}
and found that the enviroment slowed the nonadiabatic decay from 0.22~ps to 0.5~ps.
This was mainly ascribed to the presence of hydrogen bonds between guanine and cytosine.

{
The {{\sc sharc}} dynamics code,\cite{Mai2015IJQC,Richter2011JCTC,Mai2018WCMS,sharc} developed by Gonz\'alez and coworkers, has been interfaced with many quantum chemistry programs.
{The {{\sc sharc}} code is named after the surface hopping including arbitrary couplings (SHARC) model\cite{Richter2011JCTC,Mai2015IJQC} and is capable of simulating intersystem crossing dynamics.
FSSH dynamics simulations without intersystem crossing can also be performed with {\sc sharc}.}
}

One of the earilest applications of {{\sc sharc}} is the excited-state dynamics simulation of the $\mathrm{SO}_2$ molecule.\cite{Mai2014JCP}
The MRCIS method was employed to describe the correct dissociation limit.
In this study, the simulations were performed with and without inclusion of triplet states.
With only singlet states, very few trajectories decayed to the ground state;
however, when triplet states are also considered, almost 50\% of the total excited-state singlet population was transferred to triplet states, in agreement with experimental results.\cite{Wilkinson2014JCP}
This work was followed by simulations of modified nucleobases,
specifically those with heavy atoms:
2-thiocytosine,\cite{Mai2016NatCommun} 5-azacytosine,\cite{Borin2017PCCP} and 5-bromouracil,\cite{Mai2016NatCommun} all of which were reported using MRCIS.

Another example is the study on pyrrole in 2009\cite{Vazdar2009MP}
with MRCISD and the FSSH algorithm,
in which five singlet states were included in the dynamics, describing the deformation of the molecule and the timescale of the dynamics.
Additionally, an MRCIS-QM/MM approach was used to study the PSB3 cation,\cite{Ruckenbauer2013JPCA}
showing that solvent polarity affects the crossing seam which in turn affects the decay of the excited state.

\subsection{CASPT2 Applications}

In 2007, Coe \textit{et al.}\cite{Coe2007JPCA} used MS-CASPT2 analytical nuclear gradients with partial internal contraction\cite{Celani2003JCP} to perform
\textit{ab initio} multiple spawning (AIMS)\cite{Ben-Nun2000CP} dynamics for the intramolecular proton transfer in methyl salicylate,
and demonstrated the importance of dynamical electron correlation for the accurate description of the barrier heights and timescales.
In this work,  Coe \textit{et al.} used the single-state PIC-CASPT2 nuclear gradient code
while noting that MS-CASPT2 analytical gradients and derivative couplings would be an important step forward.\cite{Coe2007JPCA} 

After development of the PIC-MS-CASPT2 nuclear gradient by Celani and Werner in the mid 2000s (which has never been published),
the first MS-CASPT2 dynamics with AIMS was demonstrated for ethylene in 2009.\cite{Tao2009JPCA}
In this study, the authors approximated the derivative coupling vectors with those obtained by the relaxed CASSCF wave functions. 
Further studies on ethylene using MS-CASPT2 uncovered the discrepancies between the excited-state life time obtained in the previous computational and experimental studies\cite{Tao2011JCP}
and explored the mechanism of non-radiative decay pathways.\cite{Allison2012JCP}
A simulation including a low-lying Rydberg state using analytical derivative couplings was later reported with findings that this state was not significantly populated.\cite{Mori2012JPCA}

The same approach has been applied to several polyenes.
In 2012, cyclopentadiene ($\mathrm{CPD}$) and its tetramethylated analog ($\mathrm{Me}_4$-$\mathrm{CPD}$) were studied,\cite{Kuhlman2012FaradayDiscuss}
in which the methyl groups in $\mathrm{Me}_4$-$\mathrm{CPD}$ were replaced by an atom with a mass of 15 a.m.u. to reduce the cost of simulation.
The AIMS simulation reproduced the slower decay in $\mathrm{Me}_4$-$\mathrm{CPD}$ in comparison to CPD, which was reported eariler experimentally.\cite{Schalk2010JPCA}
It also showed that the methyl substituents are responsible for slowing the out-of-plane motion of the ring,
a necessary motion to reach the conical intersections.  The simulated induction time, which is the time before the population decay starts, was 25~fs~for CPD and 71~fs~for $\mathrm{Me}_4$-CPD.
A similar conclusion was later reached 
for $\mathrm{Me}_6$-CPD using the same approximation, which yielded the simulated induction time of 108~fs.\cite{Wolf2014PCCP}
AIMS simulations of butadiene using MS-CASPT2 were also presented in 2018.\cite{Glover2018JCP}

In addition, AIMS simulations with MS-CASPT2 were performed for PSB3\cite{Liu2016JPCB} and sulfine.\cite{Mignolet2016ACIE}
In PSB3, the dynamical correlation was found to make the conical intersection associated with the C=N transition less accessible,
which was predicted eariler by the calculations that located the MECIs.\cite{Mori2010JCP}
Furthermore, to investigate the diverse product formation in sulfine degradation,\cite{Mignolet2016ACIE} the long-time dynamics were simulated:
AIMS simulations were performed until sulfine decayed to the ground state,
from which point the dynamics was switched to Born--Oppenheimer molecular dynamics (BOMD) with density functional theory (DFT/PBE0).
It was found that, although sulfine undergoes nonadiabatic decay along only one relevant conical intersection,
the excess energy in the ground state causes the system to reach nine different products.

Apart from these AIMS simulations, PIC-CASPT2 was also used in the Zhu--Nakamura surface hopping (ZNSH) simulation of the ring-opening reaction for cyclohexadiene,\cite{Ohta2015CP}
QM/MM ZNSH dynamics of Wolff rearrangements of 1,2,3-thiadazole in acetonitrile,\cite{Liu2017JCP}
and single-state simulations of ethylene [2+2] photocycloaddition in the triplet state.\cite{Muchova2017IJQC}

\begin{figure}[tb]
	\includegraphics[width=1.0\linewidth]{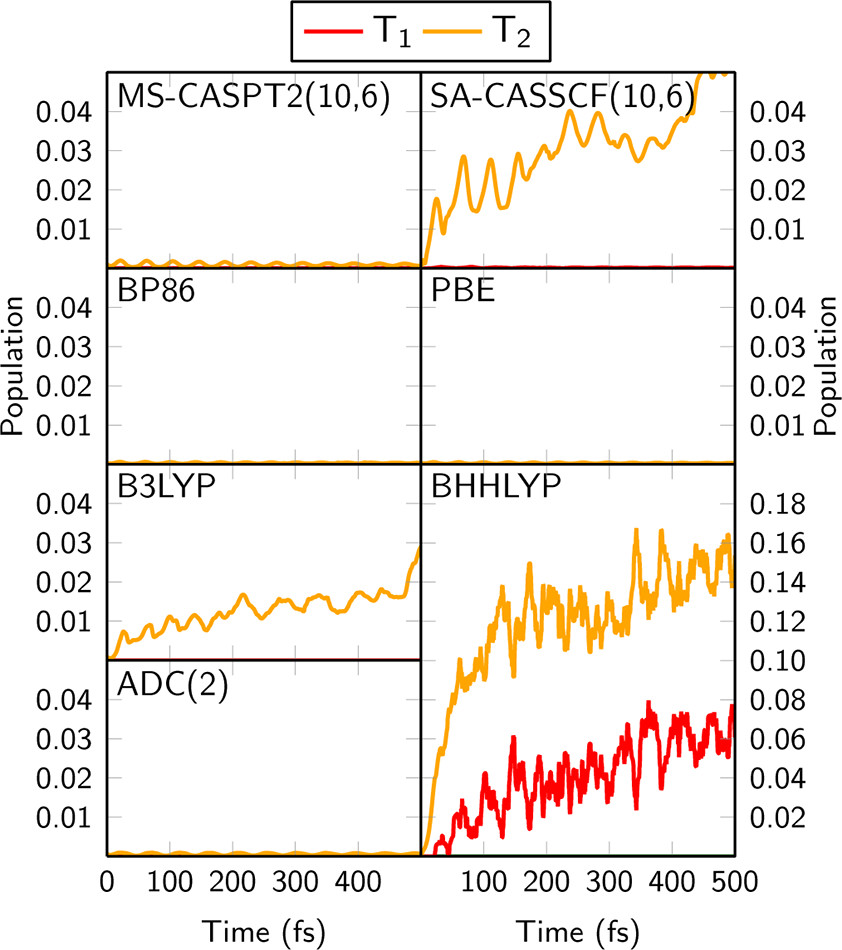}
	\caption{Dynamics population of the $\mathrm{T}_1$ (red) and $\mathrm{T}_2$ (orange) states in thioformaldehyde. In MS-CASPT2, the population transfer to the triplet state is not obtained, which is consistent with experimental results.\cite{Clouthier1983ARPC,Kawasaki1983CP,Moule2002JPCA}
Reprinted with permission from Ref.~\onlinecite{Mai2019JCTC2}. Copyright 2019 American Chemical Society.
	\label{figure:mai}}
\end{figure}

With the recent increase in available computer hardware, 
FIC-CASPT2 numerical gradients have begun to be used in surface hopping dynamics simulations, even though
the computation of numerical gradients are inherently less efficient and accurate than the analytical variant.
The mechanisms of deactivation ({mainly} involving intersystem crossing)
in the 2-thiouracil nucleobase\cite{Mai2016JPCL} and thioformaldehyde.\cite{Mai2019JCTC2}
have been studied using {the {\sc molcas}--{\sc sharc} interface together with numerical FIC-MS-CASPT2 nuclear gradients.}
In these studies, spin--orbit coupling was included using the restricted active space state interaction (RASSI) approach in {\sc molcas}.\cite{molcasref} 
The authors tested MS-CASPT2, SA-CASSCF, DFT, and ADC(2) for thioformaldehyde and found that  
MS-CASPT2 reliably reproduced the experimental observation\cite{Clouthier1983ARPC,Kawasaki1983CP,Moule2002JPCA}
that there is no intersystem crossing within a sub-picosecond time scale.
Note that the ADC(2) method predicted the intersystem crossing step as accurately as MS-CASPT2, but it overestimated the energy gap between $\mathrm{T}_2$ and $\mathrm{S}_1$ states, leading to the oscillatory amplitudes of the $\mathrm{T}_2$ population being too small
(Fig.~\ref{figure:mai}).\cite{Mai2019JCTC2} 
{Gomez {\it{et al.}} also performed MS-CASPT2 dynamics for 1,1-difluoroethylene,\cite{Gomez2019PCCP} using a minimal (2\textit{e},2\textit{o}) space and an extended (2\textit{e},6\textit{o}) active space that takes into account 3s and 3p Rydberg states (three and eleven MS-CASPT2 states were included in the calculation, respectively),
observing that the singlet Rydberg states play an important role in the population decay.}

\begin{figure*}[tb]
	\includegraphics[width=0.8\linewidth]{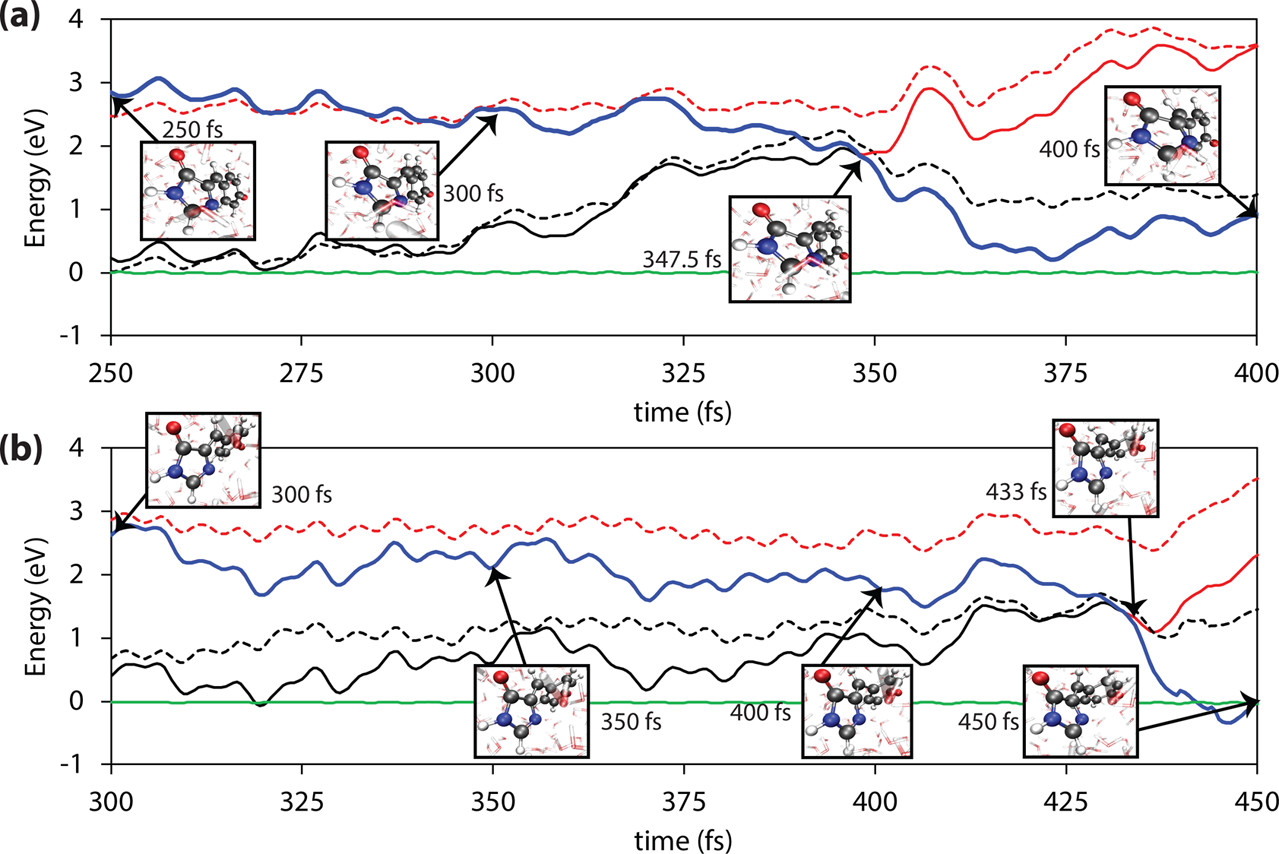}
	\caption{Dynamics trajectories with (solid) and without (dashed) solvent for the (a) $I$-channel (b) $P$-channel of $p$HBDI. 
	The decay through the $I$-channel was dominant in contrast with CASSCF results.\cite{Olsen2010JACS,Zhao2014JCP}
 Reprinted with permission from Ref.~\onlinecite{Park2017JCTC2}. Copyright 2017 American Chemical Society. \label{figure:gfpdynamics}}
\end{figure*}

One of the recent breakthroughs in this area has been the development of the analytical FIC-CASPT2 nuclear gradient program in {\sc bagel}.
FIC-CASPT2 on-the-fly surface hopping simulations have been performed using {\sc bagel}\cite{Shiozaki2018WCMS,Park2017JCTC,MacLeod2015JCP,Vlaisavljevich2016JCTC} 
together with both {\sc Newton-X}\cite{Barbatti2014WIREs,newtonx} and {\sc sharc}.\cite{Mai2018WCMS,sharc} 
We demonstrated the computational tool for QM/MM nonadiabatic decay simulations of \textit{p}HBI and adenine in water;\cite{Park2017JCTC2} 
For \textit{p}HBI (Fig.~\ref{figure:gfpdynamics}), 
the inclusion of dynamical correlation resulted in twisting channels, while previous CASSCF studies\cite{Olsen2010JACS,Zhao2014JCP} predicted otherwise.

\begin{figure}[tb]
	\includegraphics[width=0.8\linewidth]{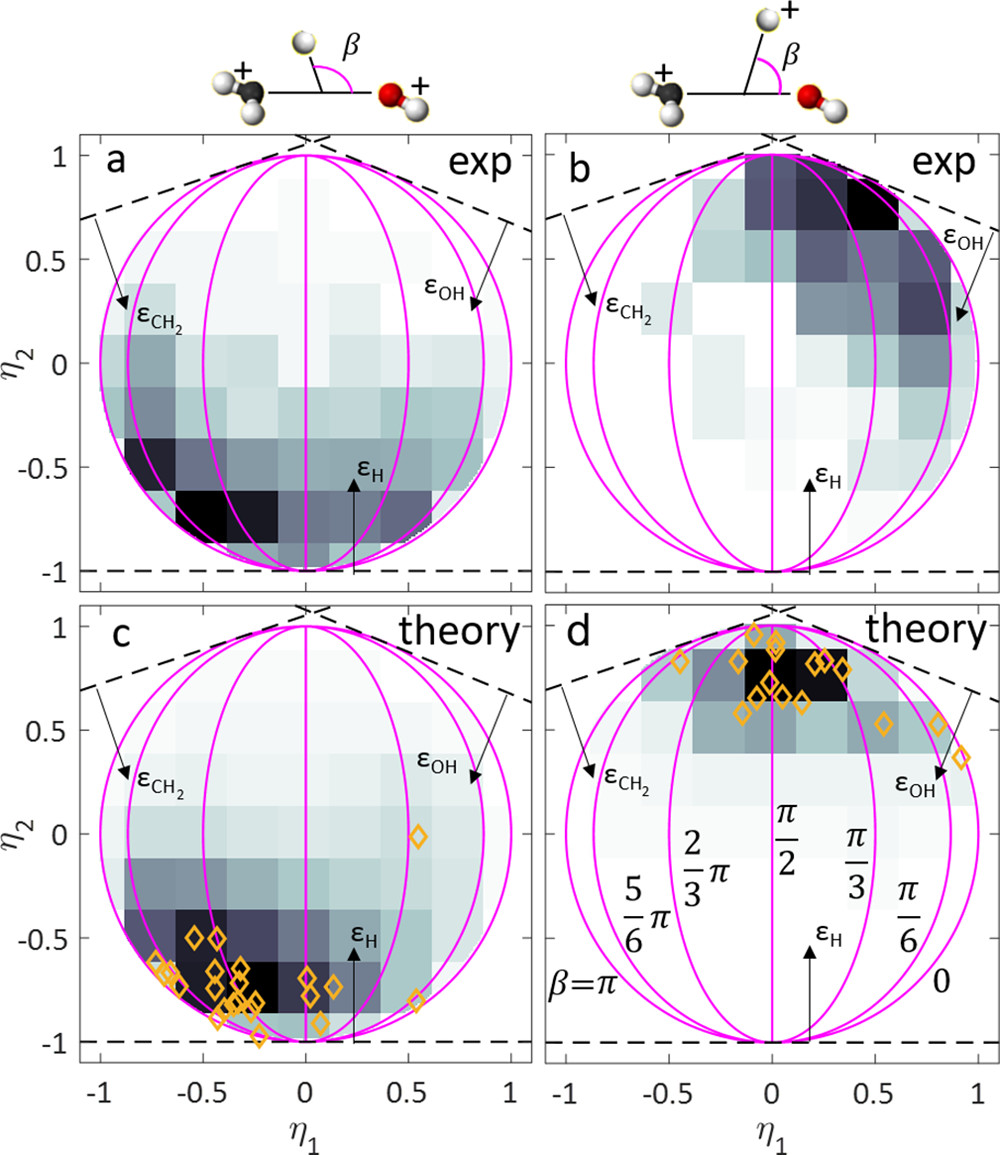}
	\caption{Experimental (top) and computed (bottom) Dalitz plots of three-body correlations for the simulations of the Coulomb explosion in methanol. 
	The diamonds show the final momentum correlations of the calculated trajectories, while the theory gray scale shows a simulated realistic experimental response.
Reprinted with permission from Ref.~\onlinecite{Luzon2019JPCL}. Copyright 2019 American Chemical Society. 		\label{figure:luzon}}
\end{figure}

The {\sc bagel}--{\sc Newton-X} interface has since been used by various users.
For example, FIC-XMS-CASPT2 FSSH dynamics simulations were used by Luzon \textit{et al.} to interpret the Coulomb explosion experiment for methanol after laser-induced multiple electron ionization.\cite{Luzon2019JPCL}
To model the process, 
two electrons were removed from the molecule at the neutral geometry, 
and the seven lowest dication states were used to initiate nonadiabatic dynamics simulations.
Overall, the FSSH simulation exhibited excellent agreement with experiment.
In particular, the correlation of three-body momentum with the ejection of a neutral H or proton (Fig.~\ref{figure:luzon}) 
suggests that these calculations are reliable
for interpreting the underlying mechanisms.

Another example is the FSSH simulations of the ring-opening reaction of 1,3-cyclohexadiene performed using XMS-CASPT2.\cite{Polyak2019JCTC}
This system is an infamous target for SA-CASSCF studies. With the full-valence (6\textit{e},6\textit{o}) active space,
the ordering of the $1^1B$ and $2^1A$ states near the Franck--Condon (FC) point is swapped with respect to the experimental observation.
Using a different active space, (6\textit{e},4\textit{o})\cite{Kim2012JPCA} or (14\textit{e},8\textit{o}),\cite{Lei2016JPhotochemPhotobiolA} the state ordering near the FC point is correct,
but accuracy at points far from the FC point, which are also sampled during the MD simulation, cannot be ensured with unbalanced selection of active space.
With the full-valence active space and XMS-CASPT2,
the correct state ordering is obtained.
Polyak {\it{et al.}} observed the ultrafast decay to the ground state (89 $\pm$ 9~fs) with large quantum yield (47 $\pm$ 8~\%) of hexatriene,
which was comparable to experimental results (80~to~170~fs lifetime\cite{Minitti2015PRL,Attar2017Science,Wolf2019NatChem} and 30\%~to~40\% quantum yield).\cite{Adachi2015JPCL}
By mapping the distribution of the hopping geometries in the simulations,
these lifetime and quantum yields are
ascribed to the extended conical intersection seam along the reaction coordinate.\cite{Nenov2010JOC}

The {\sc bagel}--{\sc sharc} interface has more recently been developed
and used to simulate pyrrole dynamics with XMS-CASPT2.\cite{Heindl2019CTC}
A time constant of 64 $\pm$ 13~fs for deactivation was obtained, 
primarily attributed to N--H bond dissociation,
which is in 
agreement with the experimental value (52 $\pm$ 12~fs) for hydrogen abstraction.\cite{Roberts2013FaradayDiscuss}
This is in contrast to
those obtained by PBE0 (166/184~fs)\cite{Barbatti2010CP} and by ADC(2) (longer than 150~fs),\cite{Sapunar2015PCCP}
supporting once more the reliability of XMS-CASPT2 dynamics simulations.

\section{Background on Electronic Structure Methods}\label{sec_electronic_structure}

Now, we are at a position to review multireference electron correlation methods and their respective nuclear gradient and derivative coupling theories
that have made the above applications possible.
We do so in two sections: first, an emphasis is placed on the electronic-structure methods in this section,
followed by the nuclear gradient theories in the subsequent section.

Generally speaking, electronic wave functions are represented as a linear combination of Slater determinants (labeled by $I$), 
\begin{equation}
| \Psi \rangle = \sum_I c_{I} | I \rangle \label{ci}
\end{equation}
If the CI-expansion of the wave function includes all possible electronic configurations, the resulting energies are exact for a given basis set.
This is the so-called full CI (FCI) approach.\cite{SzaboOstlund}
As a result of the high computational cost of evaluating the complete determinant expansion, FCI calculations can only be performed on small molecules,
even when direct diagonalization of the FCI matrix is avoided and instead an iterative procedure is used.\cite{Knowles1984CPL,Harrison1989CPL}
Therefore, approximations to the FCI expansion are essential.

\subsection{Complete Active Space Self-Consistent Field (CASSCF)}

The need to overcome the limitation of the cost of FCI led to the development of a variety of approximate theoretical models, 
in particular those that can treat systems with multiple electron configurations.
Of the multiconfigurational methods that have been developed, 
the complete active space self-consistent field (CASSCF) approach\cite{Siegbahn1980PS, Roos1980CP, Siegbahn1981JCP} has been widely used in chemical computation. 
The CASSCF wave function is a generalization of the Hartree--Fock wave function and may be viewed as an embedding scheme,
in which a subset of the total electrons (the active electrons) are treated exactly using FCI in a subset of chosen orbitals (the active orbitals) embedded in the mean-field of
other electrons.
The active electrons and orbitals define the so-called active space.
The orbitals are optimized to minimize the variational energy, resulting in the CASSCF wave function.
\begin{align}
| \Psi_P \rangle = \sum_I c_{I,P} | I \rangle \label{casstate}
\end{align}
where $P$ labels physical states.
The excited states are often simulated by means of the state-averaged CASSCF (SA-CASSCF) method,\cite{Werner1981JCP2,Docken1972JCP}
because  separate optimizations of the CASSCF orbitals for each state usually results in root flipping.
SA-CASSCF optimizes the state-averaged energy,
\begin{align}
E = \sum_P w_P E_P,
\end{align}
where $w_P$ is the weighted coefficient for individual state $P$,
which can be either static or dynamically determined.\cite{Deskevich2004JCP,Glover2014JCP,Olsen2015JCP}
Extensions of the CAS wave functions include restricted active space\cite{Olsen1988JCP,Malmqvist1990JPC} and generalized active space.\cite{Ma2011JCP}
The CASSCF wave functions are often used as a reference for multireference methods that account for dynamical correlation, which are to be discussed below. 

\subsection{Multireference Electron Correlation Methods}

In the post-CASSCF framework, dynamical correlation is incorporated by
configuration interaction (CI), perturbation (PT), and coupled cluster (CC) theories,
which we review in this section.
Several reviews are available in previous issues of \textit{Chemical Reviews} on the CI and CC theories.\cite{Lyakh2012CR,Lischka2018CR,CrespoOtero2018CR,Szalay2012CR}

\subsubsection{Configuration Interaction (CI) and Coupled Cluster (CC)}

The multireference CI approach (MRCI) is one of the viable approaches to describe dynamical correlation on the basis of CASSCF reference wave functions.\cite{Szalay2012CR,Lischka2018CR}
The uncontracted MRCI method was developed as a way to restrict the expansion space when FCI is not feasible.\cite{Buenker1974}
In this approach, the wave functions are expanded in the configurations generated by applying the excitation operator
to a number of chosen reference Slater determinants.
The uncontracted MRCI wave function is written as a linear combination of the excited determinants (or configuration state functions), as
\begin{align}
| \Psi^\mathrm{MRCI} \rangle
& = \sum_I c_I | I \rangle 
+ \sum_I \sum_{i} \sum_{a} c_{ai}^I \hat{E}_{ai} | I \rangle
\nonumber \\
&+ \sum_I \sum_{ij} \sum_{ab} c_{ai,bj}^I \hat{E}_{ai,bj} | I \rangle
+ ...
\end{align}
where $\hat{E}$ are the standard excitation operators.
As in the single-reference CI case,
truncated MRCI is not size-consistent.
This size inconsistency is partly corrected for by the Davidson correction scheme,\cite{Szalay2012CR,Langhoff1974IJQC}
or by other approximate models such as {
size-consistent multireference CI [SC-MRCI, MR(SC)$^2$CI],\cite{Malrieu1994JCP,Daudey1993JCP,Meller1996JCP,Adamowicz1996JCP,Meller2003MP}	
MR averaged coupled pair functional (MR-ACPF),\cite{Gdanitz1988CPL,Gdanitz2001IJQC}
and MR averaged quadratic coupled cluster (MR-AQCC).\cite{Szalay1993CPL,Szalay1995JCP}}

The concept of internal contraction was first reported by Meyer in the context of pair theory,\cite{Meyer1977Bookchapter}
and was later introduced in the internally contracted (IC-)MRCI method.\cite{Werner1982JCP,Werner1988JCP,Shavitt2002,Werner1987ACP}
The internal contraction greatly reduces the number of basis functions for correlated wave functions, from factorial to polynomial scaling with respect to the size of active spaces.
Since it maintains the property of exactly spanning the first-order interaction space, the errors due to internal contraction is often marginal (typically only 2--4 \% of the correlation energy).\cite{Werner1987ACP,Peterson1993JCP,Peterson1993JCP2,Angeli2012MP}
When internal contraction is used, the excited configurations are generated by applying the excitation operators to the reference wave function as a whole,
\begin{align}
| \Psi^\mathrm{ic-MRCI} \rangle
& = \sum_I c_I | I \rangle 
+ \sum_{i} \sum_{a} c_{ai} \hat{E}_{ai} | \Psi^{(0)} \rangle
\nonumber \\
&+ \sum_{ij} \sum_{ab} c_{ai,bj} \hat{E}_{ai,bj} | \Psi^{(0)} \rangle
+ ...
\end{align}
Note that the internally contracted basis, defined above, is not orthonormal and requires special care. 

There are a myriad of multireference coupled cluster (MRCC) methods,
such as the state-universal MRCC (SU-MRCC),\cite{Jeziorski1984JCP}
Brillouin--Wigner MRCC,\cite{Masik1998AdvQuantumChem,Masik1998JCP}
Mukherjee MRCC,\cite{Mahapatra1999JCP}
and H2E2-MRCC.\cite{Kong2009IJQC}
In particular, the works by the groups in Mainz\cite{Evangelista2011JCP,Hanauer2011JCP}
have introduced the internal contraction concept to MRCC. 
Each MRCC model has strength and weakness; MRCC is still under active development.

\subsubsection{Perturbation Theories (PT)}\label{sec_caspt2}

Perturbation theory is another avenue to account for dynamical correlation in the multireference framework.
There have been many attempts to generalize the single-reference case of M{\o}ller--Plesset second-order perturbation theory (MP2) to multireference theories;
for historical accounts, see an excellent perspective by Pulay.\cite{Pulay2011IJQC}
Uncontracted multireference perturbation theories (MRPT) include the implementation by Murphy and Messmer,\cite{Murphy1991CPL,Murphy1992JCP}
the method by Kozlowski and Davidson,\cite{Kozlowski1994JCP}
MRMP2 by Hirao,\cite{Hirao1992CPL,Hirao1992CPL2,Hirao1993CPL}
and uncontracted $n$-electron valence state perturbation theory (NEVPT2).\cite{Angeli2001JCP,Angeli2002JCP}
Interestingly, internally contracted multireferene perturbation theory predates these uncontracted theories; 
the first such approach was proposed by Roos {\it{et al.}} in 1982.\cite{Roos1982CP} 
Subsequently, Pulay and coworkers proposed the generalized MP2 (GMP2) method,
which was demonstrated with a two-configuration reference function.\cite{Wolinski1987CPL,Wolinski1989JCP}

Among others, complete active space second-order perturbation theory (CASPT2) by Roos and coworkers\cite{Andersson1990JPC,Andersson1992JCP} is the most widely used approach to date.
The CASPT2 method is a generalization of GMP2 for CAS reference functions. 
In the original formulation,\cite{Andersson1990JPC,Andersson1992JCP} internal contraction is used for all of the basis (hereafter referred to as FIC-CASPT2),
\begin{align}
| \Psi^{(1)} \rangle = \sum_{\Omega} T_\Omega \hat{E}_\Omega | \Phi_0 \rangle = \sum_\Omega T_\Omega | \Omega \rangle \equiv \hat{T}  | \Phi_0 \rangle,
\end{align}
where the operator $\hat{E}_\Omega$ is the standard two-electron excitation operator (including the so-called semi-internal excitations), 
and $T_\Omega$ are the corresponding amplitudes.
These amplitudes are to be determined by the stationary condition of the Hylleraas functional, which is 
\begin{align}
\sum_{\Omega'} \langle \Omega | \hat{H}^{(0)} - E^{(0)} | \Omega' \rangle T_{\Omega'} + \langle \Omega | \hat{H} | \Phi_0 \rangle = 0.\label{ampeq}
\end{align}
The definition of the zeroth-order Hamiltonian in CASPT2 is
\begin{align}
\hat{H}^{(0)} = \hat{P} \hat{f} \hat{P} + \hat{Q} \hat{f} \hat{Q},
\end{align}
where $\hat{P}$ and $\hat{Q}$ are the projectors onto the reference and first-order interacting space, respectively,
and $\hat{f}$ is the one-electron Fock operator.
The NEVPT2 method was developed by using the two-electron Dyall's Hamiltonian in the active space\cite{Dyall1995JCP} instead of the Fock operator.
The method using internally contracted functions is referred to as partially contracted (PC)-NEVPT2 theory.\cite{Angeli2001JCP,Angeli2002JCP}
In strongly contracted (SC)-NEVPT2 theory,\cite{Angeli2001JCP,Angeli2002JCP}
the configurations are further contracted with the excitation operators
so that the expansion basis has only one function within the excitation subspace.
The strong contraction has also been used in conjunction with the MRCI method.\cite{Sivalingam2016JCP}

Full internal contraction in the CASPT2 theory requires four-particle reduced density matrices (RDMs) in the active space (and diagonalization of three-particle RDMs),
which used to be a computational bottleneck.
Werner and coworkers developed a method with partial internal contraction, called RS2, to avoid this bottleneck, a scheme that is similar to Werner's MRCI internal contraction scheme.\cite{Werner1996MP}
We hereafter call it PIC-CASPT2.
In this scheme, only doubly external configurations
are internally contracted;
the first-order wave function is written as
\begin{align}
| \Psi^{(1)} \rangle = \sum_I t_I | I \rangle 
+ \sum_S \sum_a t_a^S | S^a \rangle
+ \sum_{ij} \sum_{ab} T_{ai,bj}\hat{E}_{ai,bj} | \Phi_0 \rangle,
\end{align}
where $|S^a\rangle$ denotes the singly external configuration state functions.
This contraction scheme requires up to three-particle RDMs in the active space, although the number of parameters now scales combinatorially (in contrast to full internal contraction). 

Calculating electronically excited states is one of the most important applications of multireference theory.\cite{Lischka2018CR}
Multistate (MS) perturbation theories were developed to calculate avoided crossings and conical intersections within the framework of MRPT.
Examples include multiconfigurational quasidegenerate perturbation theory [(X)MCQDPT] theory,\cite{Nakano1993JCP,Granovsky2011JCP} which is the multistate version of MRMP2,
multistate CASPT2 [(X)MS-CASPT2] theory,\cite{Finley1998CPL,Granovsky2011JCP,Shiozaki2011JCP3}
and quasidegenerate NEVPT2 (QD-NEVPT2) theory, which is the multistate version of NEVPT2.\cite{Angeli2004JCP}
The original MS-CASPT2 theory\cite{Finley1998CPL} is based on the Zaitsevskii and Malrieu's multipartitioning technique.\cite{Zaitsevskii1995CPL}
The perturbation expansion is performed using the state-specific first-order interacting space,
in which the first-order correction to the wave function is
\begin{align}
| \Psi^{(1)}_L \rangle = \hat{T}_L | L \rangle.
\end{align}
The zeroth-order Hamiltonian is
\begin{align}
\hat{H}^{(0)}_M = | M \rangle \langle M | \hat{f}^\mathrm{M} | M \rangle \langle M | + \hat{Q}^M\hat{f}^M\hat{Q}^M,\label{sssrmscaspt2}
\end{align}
where $\hat{Q}^M$ is the first-order interacting space generated by applying the excitation operators to the state $M$,
and $\hat{f}^M$ is either the state-specific Fock operator for state $M$ or the state-averaged Fock operator, depending on the implementation. 
The amplitude equation for each state is solved to obtained the first-order amplitudes,
using which the effective Hamiltonian is constructed as
\begin{align}
& H^\mathrm{eff}_{LL^\prime} = \langle L | \hat{H} | L^\prime \rangle + \langle L | \hat{T}^\dagger_{L} \hat{H} | L^\prime \rangle.
\end{align}
This effective Hamiltonian is diagonalized to arrive at the MS-CASPT2 state,
\begin{align}
\sum_{MN} H^\mathrm{eff}_{MN} R_{NP} = R_{MP} E^\mathrm{MS-CASPT2}_P.
\end{align}
The wave function of a MS-CASPT2 state is
\begin{align}
| \Psi_P \rangle = \sum_M R_{MP} \left( |M \rangle + | \Psi_M^{(1)} \rangle \right).
\end{align}
Werner and coworkers also implemented a variant that uses the union of reference states to generate the internally contracted basis in {\sc molpro}.\cite{MOLPRO}
In the MS-MR contraction scheme,
the first-order interacting space is formed from all the reference states, as
\begin{align}
| \Psi^{(1)}_L \rangle = \sum_M \hat{T}_{LM} |M \rangle = \sum_M \sum_\Omega  T_{LM,\Omega} \hat{E}_\Omega | M \rangle.
\end{align}
and the state-averaged Fock operator is used in the zeroth-order Hamiltonian, as
\begin{align}
\hat{H}^{(0)} =  \sum_M | M \rangle \langle M | \hat{f}^\mathrm{SA} | M \rangle \langle M | + \hat{Q} \hat{f}^\mathrm{SA} \hat{Q},\label{msmrmscaspt2}
\end{align}
where $\hat{f}^\mathrm{SA}$ is the Fock operator evaluated with the state-averaged electronic density.
Note that the projection onto the reference space in the first term
only includes the diagonal term to ensure that the CASSCF wave function is the eigenfunction of $\hat{H}^{(0)}$.
As pointed out in the ingenious work by Granovsky,\cite{Granovsky2011JCP} the diagonal projection in the first term leads to 
irregularity in the potential energy surfaces;
he further suggested the invariant form of the zeroth-order Hamiltonian with respect to rotation of reference functions.
In this scheme, the off-diagonal elements are included by defining rotated reference states,
\begin{align}
\hat{H}^{(0)} = \hat{P}\hat{f}^\mathrm{SA}\hat{P} + \hat{Q} \hat{f}^\mathrm{SA} \hat{Q},
\end{align}
in which the $\hat{P}$ is the complementary orthogonal projector of $\hat{Q}$.
The first term is rewritten in the same form as Eq.\eqref{msmrmscaspt2} by diagonalizing the Fock operator in the reference space, 
\begin{align}
\sum_M \langle L | \hat{f}^\mathrm{SA} | M \rangle U_{MN} = U_{LN} \tilde{E}_N^{(0)}.
\end{align}
Granovsky named the uncontracted variant XMCQDPT2,
and following the same naming convention, we call the internally contracted variant XMS-CASPT2.\cite{Shiozaki2011JCP3}

In MRMP2, (X)MCQDPT, and [(X)MS-]CASPT2, one often encounters the so-called intruder-state problem that causes spurious divergences in computed energies. 
The problem is easily recognized: if the basis functions in Eq.~\eqref{ampeq} diagonalize the zeroth-order Hamiltonian and are orthonormal,
the amplitude is formally
\begin{align}
T_\Omega = -\frac {\langle \Omega | \hat{H} | \Phi_0 \rangle}{ \langle \Omega | \hat{H}^{(0)} | \Omega \rangle - E^{(0)} }.\label{amp}
\end{align}
When the eigenvalue of the zeroth-order Hamiltonian for the basis function $\Omega$ is close to that of the reference function,
the denominator becomes close to zero,
and the amplitude becomes infinity, hence a divergent CASPT2 energy.
Note the (QD-)NEVPT2 method is known to be usually immune to this problem,
because it uses Dyall's Hamiltonian,\cite{Dyall1995JCP} which is a two-electron Hamiltonian in the active space.
To circumvent this problem within the CASPT2 framework, schemes have been proposed in which one adds a real\cite{Roos1995CPL} or imaginary\cite{Forsberg1997CPL} level shift to the denominator in Eq.~\eqref{amp}.
This is equivalent to adding the level shift to the zeroth-order Hamiltonian as
\begin{align}
\sum_{\Omega'} \langle \Omega | \hat{H}^{(0)} - E^{(0)} + \epsilon | \Omega' \rangle T_{\Omega'} + \langle \Omega | \hat{H} | \Phi_0 \rangle = 0.
\end{align}
For the imaginary shift, a trick is introduced to avoid the complex algebra throughout the energy evaluation. 
The errors due to the level shift are corrected using the first-order perturbation theory, often referred to as shift corrections.
For (X)MCQDPT, the intruder state avoidance (ISA) technique is used to remove the intruder state,\cite{Witek2002JCC}
though shift corrections are usually not included.

To improve the accuracy of CASPT2, other definitions of zeroth-order Hamiltonian have been introduced for the CASPT2 method.\cite{Ghigo2004CPL,Andersson1995TCA}
A widely known empirical modification to the zeroth-order Hamiltonian is the so-called ionization potential electron affinity (IPEA) shift.\cite{Ghigo2004CPL}
In this scheme, one assumes that the diagonal elements of the generalized Fock operator for the active orbitals
is an interpolation between the diagonal elements of the closed and virtual orbitals, as
\begin{align}
f_{pp} = -\frac{1}{2} \left[ \gamma_{pp} (\mathrm{IP})_p + (2 - \gamma_{pp}) (\mathrm{EA})_p \right].
\end{align}
The IPEA shift modifies these diagonal elements, so that the diagonal elements of the Fock matrix are
\begin{align}
f_{pp}^\mathrm{out} = -(\mathrm{IP})_p,
\end{align}
when an electron is excited \textit{out of} the orbital $p$, and
\begin{align}
f_{pp}^\mathrm{in} = -(\mathrm{EA})_p,
\end{align}
when an electron is excited \textit{into} the orbital $p$.
This modification is formally written as
\begin{align}
\hat{H}^{(0)'} = \hat{H}^{(0)} + \frac{1}{2} \epsilon \kappa_\Omega,
\end{align}
where $\kappa_\Omega$ is different for each excitation subspace (see Ref.~\onlinecite{Kepenekian2009JCP} for its explicit form)
and $\epsilon$ is referred to as the IPEA shift parameter.
In the reference software for CASPT2, {\sc molcas},\cite{molcasref} the IPEA shift parameter of 0.25 is default.
There have been many suggestions for IPEA shift parameter values:
0.25~$E_\mathrm{h}$,\cite{Ghigo2004CPL,Rudavskyi2014JCP,Kepenekian2009JCC}
0.50~$E_\mathrm{h}$,\cite{Suaud2009JACS,Vela2016JCC}
or even higher values.\cite{Kepenekian2009JCP,LawsonDaku2012JCTC}
Recently, Zobel \textit{et al.} suggested using the zero IPEA shift value for organic molecules.\cite{Zobel2017ChemSci} 

\section{Nuclear gradients and derivative coupling theory}\label{sec_gradient_theory}

In Secs.~\ref{sec_application_geomopt} and \ref{sec_application_dynamics}, we have reviewed the applications of multireference electron correlation theories
for the study of chemical systems.
In principle, it is always possible to use finite difference gradients and derivative couplings in such calculations.
The two-point finite difference gradient toward the direction of an arbitrary vector $\mathbf{x}$ is computed as
\begin{align}
\frac{\mathrm{d}E}{\mathrm{d} x} = \frac{E(\mathbf{R}+\mathbf{x}\Delta ) - E(\mathbf{R}-\mathbf{x} \Delta)}{ 2 \Delta},
\end{align}
where $\mathbf{R}$ is the current molecular geometry and
$\Delta$ is the length of the displacement.
For molecular dynamics (MD) simulations and geometry optimizations,
$\mathbf{x}$ corresponds to the Cartesian coordinates of all atoms.
In other words, the energy should be repeatedly calculated for $6 N_\mathrm{atom}$ geometric variables.
Although the finite difference gradient can be embarrassingly parallelized
and has been used in some of the aforementioned studies,
quantum chemical methods become more useful when the analytical gradient algorithms are available owing to the computational savings.
In this section, we review analytical nuclear gradients and derivative coupling theories in detail.

\subsection{Fundamentals of nuclear gradient theory}
For historical accounts of the early developments of nuclear gradient theory,
readers are referred to Ref.~\onlinecite{Pulay2014WIREs} by Pulay.
For exact wave functions at the complete basis set limit, the Hellmann--Feynman theorem holds,\cite{Hellmannbookchapter,Feynman1939PR}
\begin{align}
\frac{\mathrm{d}E}{\mathrm{d}\mathbf{R}} = \langle \Psi | \frac{\partial \hat{H}}{\partial \mathbf{R}} | \Psi \rangle.
\end{align}
This is because the derivative of wave functions, $\mathrm{d}\Psi/\mathrm{d} \mathbf{R}$, is zero due to the variational principle.
The Hellmann--Feynman theorem, however, is not applicable to virtually any quantum chemical method, because (1) wave functions are represented using finite basis functions
and (2) wave functions are often not optimized variationally. The only exceptions are mean-field models in a plane-wave or finite-element basis.

\subsubsection{Hartree--Fock gradient}

Brato{\v{z}},\cite{Bratoz1958ColloqIntCNRS} Gerratt and Mills,\cite{Gerratt1968JCP} and Pulay\cite{Pulay1969MP} independently derived, by explicit differentiation, the expression for the nuclear gradient of the Hartree--Fock method: 
\begin{align}
\frac{\mathrm{d}E}{\mathrm{d}\mathbf{R}} = \sum_{\mu\nu} d_{\mu\nu} h^\mathbf{R}_{\mu\nu} + \frac{1}{2} \sum_{\mu\nu\rho\lambda} D_{\mu\nu,\rho\lambda} (\mu\nu|\rho\lambda)^\mathbf{R}
- \sum_{\mu\nu} X_{\mu\nu} S^\mathbf{R}_{\mu\nu} \label{hfderiv}
\end{align}
where molecular integrals with superscripts $\mathbf{R}$ are the derivative integrals:
\begin{align}
&h^\mathbf{R}_{\mu\nu} = \frac{\partial h_{\mu\nu}}{\partial \mathbf{R}},\\
&S^\mathbf{R}_{\mu\nu} = \frac{\partial S_{\mu\nu}}{\partial \mathbf{R}},\\
&(\mu\nu|\rho\lambda)^\mathbf{R} = \frac{\partial (\mu\nu|\rho\lambda)}{\partial \mathbf{R}}.
\end{align} 
We often refer to the last term as the Pulay force to recognize his seminal contributions. 
The matrix that is multiplied by $S^\mathbf{R}$ is an energy-weighted density matrix,
\begin{align}
\mathbf{X} = 2 \mathbf{C}_\mathrm{occ} \boldsymbol{\epsilon} \mathbf{C}_\mathrm{occ}^\dagger 
\end{align}
and $\mathbf{C}_\mathrm{occ}$ is a slice of the MO coefficient matrix that corresponds to occupied orbitals.
The same formulas were independently derived by McIver and Komornicki for semi-empirical methods a few years later.\cite{McIver1971CPL}
Since the density matrices and energy-weighted density matrix are readily available from Hartree--Fock calculations,
the evaluation of nuclear gradients is typically less computationally demanding than the energy evaluation in this case.

\subsubsection{Coupled-Perturbed Hartree--Fock and the $Z$-vector Equation}

The first derivatives of $\mathbf{C}$ with respect to nuclear displacements are required to compute the nuclear gradients for correlated methods, 
because the energy is not stationary with respect to orbital rotations any longer.
Since the MO basis is orthonormal, 
any change in the MO coefficients can be written in term of
the unitary transformation matrix $\mathbf{U}$, 
\begin{align}
\mathbf{C} = \mathbf{C}^{0}\mathbf{U},
\end{align}
and it is convenient to use the generator of $\mathbf{U}$ as fundamental variables instead of the elements of $\mathbf{C}$.
The nuclear gradients for correlated methods are generally written as\cite{Pople1979IJQC,Brooks1980JCP}
\begin{align}
\frac{\mathrm{d}E}{\mathrm{d}\mathbf{R}}& = \sum_{\mu\nu} d_{\mu\nu} h^\mathbf{R}_{\mu\nu} + \frac{1}{2} \sum_{\mu\nu\rho\lambda} D_{\mu\nu,\rho\lambda} (\mu\nu|\rho\lambda)^\mathbf{R} \nonumber \\
& + \frac{1}{2} \sum_{xy} \left(Y_{xy} - Y_{yx} \right) U_{xy}^\mathbf{R} - \frac{1}{2}\sum_{xy}  Y_{xy} S_{xy}^\mathbf{R},\label{correlderiv}
\end{align}
where $Y_{xy}$ corresponds to the gradient of the energy with respect to the variation of $U_{xy}$, and
\begin{align}
S_{xy}^\mathbf{R} = \sum_{\mu\nu} C_{\mu x} S_{\mu\nu}^\mathbf{R} C_{\nu y}.
\end{align}
The MO orthonormality relation
\begin{align}
\mathbf{U}^{\mathbf{R}\dagger} + \mathbf{S}^\mathbf{R} + \mathbf{U}^\mathbf{R} = \mathbf{0}\label{orthogonality}
\end{align}
is explicitly considered.
The coupled--perturbed Hartree--Fock (CPHF) equation is solved to calculate $\mathbf{U}^\mathbf{R}$.
This equation has the form of
\begin{align}
\mathbf{A} \mathbf{U}^{\mathbf{R}} = \mathbf{B}^{\mathbf{R}},
\end{align}
which is a coupled Hartree--Fock equation with respect to an external perturbation.
Historically, the CPHF equation was first derived in 1960s to calculate several molecular properties.\cite{McWeeny1960RMP,McWeeny1962PR,Stevens1963JCP}
It was realized that that the changes in $\mathbf{R}$ can also be treated as a perturbation by adding the derivative integrals.\cite{Gerratt1968JCP,Caves1969JCP}
Pople and coworkers developed an efficient method for solving CPHF to calculate the Hessian of the Hartree--Fock energy,
and extended analytical gradient theory to
second-order M{\o}ller--Plesset perturbation theory (MP2).\cite{Pople1979IJQC}
The nuclear gradient of the SCF-based configuration interaction (CI) appeared
almost immediately by Schaefer and coworkers\cite{Brooks1980JCP} and Pople and coworkers.\cite{Krishnan1980JCP}

The CPHF equation should be solved for $3N_{\mathrm{atom}}$ degrees of freedom in order to calculate $( \mathbf{Y} - \mathbf{Y}^\dagger )^\dagger \mathbf{U}^\mathbf{R}$ in Eq.~\eqref{correlderiv}.
In 1984, Handy and Schaefer reported\cite{Handy1984JCP}
that this term is rearranged as
\begin{align}
( \mathbf{Y} - \mathbf{Y}^\dagger )^\dagger \mathbf{U}^\mathbf{R} = ( \mathbf{Y} - \mathbf{Y}^\dagger )^\dagger \mathbf{A}^{-1} \mathbf{B}^\mathbf{R} \equiv \mathbf{Z}^{\dagger} \mathbf{B}^\mathbf{R}.\label{handyshaefer}
\end{align}
This means that the same contribution is computed by only solving one linear equation,
\begin{align}
\mathbf{A}^\dagger \mathbf{Z} = \mathbf{Y} - \mathbf{Y}^\dagger.\label{srzvector}
\end{align}
This equation is the $Z$-vector equation.
It was originally applied to the CI nuclear gradient,\cite{Handy1984JCP}
and shortly thereafter to MP2\cite{Pulay1986TCA,Handy1985CPL}
and coupled-cluster (CC) nuclear gradients.\cite{Adamowicz1984IJQC,Hoffmann1987CPL,Scuseria1987IJQC}

\subsubsection{Relaxed (Effective) Density}

In the course of early developments,
transformation of the integral derivative from the AO to MO basis was a bottleneck for 
calculating the gradients,\cite{Rice1985CPL}
since it requires $3N_{\mathrm{atom}}$ integral transformations.
After the development of the $Z$-vector equation, it turned out that the MO derivative integral is not needed at all
(i.e., only one transformation of an MO quantity to the AO basis is needed).
By forming the density-matrix-like quantities in the AO basis,\cite{Rice1985CPL,Salter1987CPL}
the gradient is written as
\begin{align}
\frac{\mathrm{d}E}{\mathrm{d}\mathbf{R}} = \sum_{\mu\nu} \tilde{d}_{\mu\nu} h_{\mu\nu}^\mathbf{R} + \frac{1}{2} \sum_{\mu\nu,\rho\lambda} \tilde{D}_{\mu\nu,\rho\lambda} (\mu\nu | \rho\lambda)^\mathbf{R} + \sum_{\mu\nu} \tilde{X}_{\mu\nu}S_{\mu\nu}^\mathbf{R}\label{effectivedensity}
\end{align}
These quantities, $\tilde{\mathbf{d}}$ and $\tilde{\mathbf{D}}$ are now widely referred to as the relaxed or effective densities.
Several important properties of molecules, including dipole moments,
are conventionally computed using these densities. 
The formulation of the relaxed density matrix led to the direct algorithm, which does not store any integral derivatives
and computes them on the fly.\cite{Frisch1990CPL}

\subsection{Early Developments for Multireference Methods}

Up to this point, we have discussed single reference methods that use one Slater determinant as the reference wave function.
For further reviews on single-reference methods, readers should see these excellent reviews.\cite{Stanton2000IntRevPhysChem,Pulay2014WIREs,Pulay1995bookchapter}
Now, we focus on early developments of nuclear gradient theory for multireference methods, the main subject of this review.

\subsubsection{CASSCF Gradient}

The CASSCF gradient has a form similar that of Hartree--Fock,\cite{Kato1979CPL,Osamura1982JCP,Dupuis1981JCP,Page1984JCP,Pulay1977bookchapter,Kumanova1972MolPhys} as
\begin{align}
\frac{\mathrm{d}E}{\mathrm{d}\mathbf{R}} & = \sum_{\mu\nu} d_{\mu\nu} h_{\mu\nu}^\mathbf{R} + \frac{1}{2} \sum_{\mu\nu\rho\lambda} D_{\mu\nu,\rho\lambda} (\mu\nu|\rho\lambda)^\mathbf{R} - \sum_{\mu\nu} {X}_{\mu\nu} S_{\mu\nu}^\mathbf{R}.
\end{align}
The generalized energy-weighted density matrix is
\begin{align}
&X_{ij}^\mathrm{MO} = h_{ij} d_{ij} + \sum_{kl}D_{ij,kl} (ij|kl),\nonumber\\
&\mathbf{X}^\mathrm{AO} = \mathbf{C}_\mathrm{occ} \mathbf{X}^\mathrm{MO} \mathbf{C}^\dagger_\mathrm{occ}.
\end{align}
This reduces to the Hartree--Fock gradient when there are no active orbitals.\cite{Pulay1977bookchapter}

\subsubsection{Coupled-perturbed CASSCF and $Z$-vector equation}

The energy is no longer optimized with respect to the orbital coefficients in SA-CASSCF.
The nuclear gradient of the SA-CASSCF state $P$ is
\begin{align}
\frac{\mathrm{d}E_P}{\mathrm{d}\mathbf{R}} & = \sum_{\mu\nu} d_{\mu\nu}^{PP} h_{\mu\nu}^\mathbf{R} + \frac{1}{2} \sum_{\mu\nu\rho\lambda} D_{\mu\nu,\rho\lambda}^{PP} (\mu\nu|\rho\lambda)^\mathbf{R} \nonumber \\
& + \frac{1}{2}\sum_{xy} \left( Y_{xy} - Y_{yx} \right) U_{xy}^\mathbf{R}  - \frac{1}{2}\sum_{xy}  Y_{xy} S_{xy}^\mathbf{R},
\end{align}
where
\begin{align}
Y_{xy} = 4 \left( \sum_z d_{yz}^{PP} h_{xz} + \sum_{zwu} D_{yz,wu}^{PP} (xz|wu) \right),\label{orbderivsacasscf}
\end{align}
and
\begin{align}
& d_{rs}^{PP} = \langle \Psi_P | \hat{E}_{rs} | \Psi_P \rangle, \\
& D_{rs,tu}^{PP} = \langle \Psi_Q | \hat{E}_{rs,tu} | \Psi_P \rangle.
\end{align}
The orthogonality relation [Eq.~\eqref{orthogonality}] is explicitly taken into account.
The orbital derivatives $\mathbf{U}^\mathbf{R}$ can be obtained by the coupled-perturbed multiconfigurational self-consistent field (CPMCSCF) equation,
which is the multireference extension to the CPHF equation.\cite{Osamura1982JCP}
The CPMCSCF equation has the form of
\begin{align}
 \begin{pmatrix} \mathbf{A}_{UU} & \mathbf{A}_{cU}^\dagger \\[8pt] \mathbf{A}_{cU} & \mathbf{A}_{cc} \end{pmatrix} 
\begin{pmatrix} \mathbf{U}^{\mathbf{R}} \\[8pt] \mathbf{c}^{\mathbf{R}} \end{pmatrix}  =
\begin{pmatrix} \mathbf{B}_U^\mathbf{R} \\[8pt] \mathbf{B}_c^\mathbf{R}  \end{pmatrix}.
\end{align}
If there is only one configuration in the reference function, this reduces to the CPHF equation.
This equation was first used for the MRCI nuclear gradient,\cite{Osamura1982JCP}
before it was applied to SA-CASSCF.

The CPMCSCF equation is rewritten in the $Z$-vector form as in the case of the CPHF equation.
A simple derivation was suggested by Handy and Schaefer.\cite{Handy1984JCP}
Following the derivation of the $Z$-vector equation in the CPHF equation yields
\begin{align}
\begin{pmatrix} \mathbf{Y}_U - \mathbf{Y}_U^\dagger \\[8pt] \mathbf{y}_c \end{pmatrix}^\dagger
\begin{pmatrix} \mathbf{U}^\mathbf{R} \\[8pt] \mathbf{c}^\mathbf{R} \end{pmatrix}
& =
\begin{pmatrix} \mathbf{Y}_U - \mathbf{Y}_U^\dagger \\[8pt] \mathbf{y}_c \end{pmatrix}^\dagger
 \begin{pmatrix} \mathbf{A}_{UU} & \mathbf{A}_{cU}^\dagger \\[8pt] \mathbf{A}_{cU} & \mathbf{A}_{cc} \end{pmatrix}^{-1}
 \begin{pmatrix} \mathbf{B}_U^\mathbf{R} \\[8pt] \mathbf{B}_c^\mathbf{R}  \end{pmatrix} \nonumber \\
& =
\begin{pmatrix} \mathbf{Z}_U \\[8pt] \mathbf{z}_c \end{pmatrix}^\dagger
\begin{pmatrix} \mathbf{B}_U^\mathbf{R} \\[8pt] \mathbf{B}_c^\mathbf{R}  \end{pmatrix}
\end{align}
and therefore,
\begin{align}
\begin{pmatrix} \mathbf{A}_{UU} & \mathbf{A}_{cU}^\dagger \\[8pt] \mathbf{A}_{cU} & \mathbf{A}_{cc} \end{pmatrix}^{\dagger}
\begin{pmatrix} \mathbf{Z}_U \\[8pt] \mathbf{z}_c \end{pmatrix}
= 
\begin{pmatrix} \mathbf{Y}_U - \mathbf{Y}_U^\dagger \\[8pt] \mathbf{y}_c \end{pmatrix}.\label{zcasscf_direct}
\end{align}
This is the $Z$-vector equation for the multireference wave function.
Although $\mathbf{y}_c = \mathbf{0}$ in the SA-CASSCF case,
$\mathbf{z}_c \neq \mathbf{0}$,
and $\mathbf{B}^\mathbf{R}_c$ should be taken account in the gradient evaluation.

Note, however, that SA-CASSCF nuclear graidents can be more efficiently computed by a semi-numerical algorithm
\begin{align}
\frac{\mathrm{d} E_P}{\mathrm {d} \mathbf{R}} & =
\sum_{\mu \nu} \frac{\mathrm{d} d_{\mu \nu}^\mathrm{SA}}{\mathrm {d} W_P} h_{\mu\nu}^\mathbf{R} +
\frac{1}{2} \sum_{\mu\nu\rho\lambda} \frac{\mathrm{d} D_{\mu \nu, \rho \lambda}^\mathrm{SA}}{\mathrm {d} W_P} (\mu \nu | \rho \lambda)^\mathbf{R} \nonumber \\
& - \sum_{\mu\nu} \frac{\mathrm{d} X_{\mu\nu}^\mathrm{SA}}{\mathrm {d} W_P} S_{\mu \nu}^\mathbf{X}.
\end{align}
in which a finite difference formula with respect to the state weights is used to obtain the SA-CASSCF nuclear gradient.\cite{Granovsky2015JCP}

\subsubsection{SA-CASSCF derivative coupling}

Derivative coupling is a quantity that describes the coupling between the nuclear and electronic motions,
and is a central quantity for describing nonadiabatic phenomena within the Born--Oppenheimer approximation.
For the adiabatic states (which are obtained within the Born--Oppenheimer approximation) $P$ and $Q$, 
the derivative coupling is
\begin{align}
\mathbf{h}^{QP} = \langle \Psi_Q | \frac{\mathrm{d}}{\mathrm{d}\mathbf{R}} | \Psi_P \rangle.
\end{align}
By inserting Eq.~\eqref{casstate}, we get
\begin{align}
& \mathbf{h}^{QP} = \mathbf{h}_{\mathrm{CI}}^{QP} + \mathbf{h}_{\mathrm{det}}^{QP}, \\
& \mathbf{h}_{\mathrm{CI}}^{QP} = \sum_I c_{I,Q}\frac{\mathrm{d}c_{I,P}}{\mathrm{d}\mathbf{R}}, \\
& \mathbf{h}_{\mathrm{det}}^{QP} = \sum_{JI} c_{J,Q} c_{I,P} \braket{ J }{ \frac{\mathrm{d}I}{\mathrm{d}\mathbf{R}}}.
\end{align}
The terms $\mathbf{h}_\mathrm{CI}$ and $\mathbf{h}_\mathrm{det}$ are conventionally called the CI and determinant terms, respectively.\cite{Park2017JCTC,Lengsfield1984JCP,Lengsfield1992ACP,Galvan2016JCTC,Lischka2004JCP,Chernyak2000JCP}
The CI term here is also referred to as the interstate coupling.
It is one of the vectors that defines the branching plane,
which plays an important role in dynamics at conical intersections (the other is the gradient difference vector).\cite{Matsika2011ARPC,Bearpark1994CPL,Manaa1993JCP,Mori2009CPL}

The CI term is calculated by
\begin{align}
\mathbf{h}_{\mathrm{CI}}^{QP} = \frac{1}{E_P - E_Q} \sum_{IJ} c_{I,Q} \frac{\mathrm{d}H_{IJ}}{\mathrm{d}\mathbf{R}} c_{J,P},
\end{align}
from the CI eigenvalue relationship, $\mathbf{H}\mathbf{c}_I = \mathbf{E} \mathbf{c}_I$.
Since this expression is similar to that for the SA-CASSCF energy,
\begin{align}
\frac{\mathrm{d}E_P}{\mathrm{d}\mathbf{R}} = \sum_{IJ} c_{I,P} \frac{\mathrm{d}H_{IJ}}{\mathrm{d}\mathbf{R}} c_{J,P},
\end{align}
the CI term is evaluated using a similar expression to that for the analytical gradient,
\begin{align}
\left( E_P - E_Q \right) \mathbf{h}_{\mathrm{CI}}^{QP} & = \sum_{\mu\nu} \bar{d}_{\mu\nu}^{QP} h^\mathbf{R}_{\mu\nu} + \frac{1}{2} \sum_{\mu\nu\rho\lambda} \bar{D}_{\mu\nu,\rho\lambda}^{QP} (\mu\nu|\rho\lambda)^\mathbf{R} \nonumber \\
& + \frac{1}{2} \sum_{xy} \left(Y_{xy} - Y_{yx} \right) U_{xy}^\mathbf{R} - \frac{1}{2}\sum_{xy}  Y_{xy} S_{xy}^\mathbf{R}.
\end{align}
Here, we used the symmetrized transition density matrix elements,
\begin{align}
& \bar{d}_{rs}^{QP} = \frac{1}{2} \langle \Psi_Q | \hat{E}_{rs} + \hat{E}_{sr} | \Psi_P \rangle, \\
& \bar{D}_{rs,tu}^{QP} = \frac{1}{2} \langle \Psi_Q | \hat{E}_{rs,tu} + \hat{E}_{tu,sr} | \Psi_P \rangle.
\end{align}
The orbital derivative is
\begin{align}
Y_{rs} = 4 \left( \sum_t \bar{d}_{st}^{QP} h_{rt} + \sum_{tuv} \bar{D}_{st,uv}^{QP} (rt|uv) \right).
\end{align}
The indices in the orbital derivative is restricted to the active indices,
because the elements of $\bar{d}$ and $\bar{D}$ are zero outside the active space.

The determinant term is calculated by rewriting the operator $\mathrm{d}/\mathrm{d}\mathbf{R}$ as a one-electron operator,\cite{Lengsfield1984JCP}
\begin{align}
\frac{\mathrm{d}}{\mathrm{d}\mathbf{R}} = \sum_{rs} \left( \mathbf{U}^\mathbf{R} + \boldsymbol{\sigma}^\mathbf{R} \right)_{rs} \hat{E}_{rs},
\end{align}
where we introduce the asymmetric overlap derivative integral
\begin{align}
\sigma^\mathbf{R}_{rs} = \sum_{\mu \nu}  C_{\mu r} C_{\nu s} \braket{\mu}{\frac{\mathrm{d}\nu}{\mathrm{d}\mathbf{R}}},
\end{align}
and the determinant term is
\begin{align}
\mathbf{h}^{QP}_\mathrm{det} = \sum_{rs} d_{rs}^{QP} \left( \mathbf{U}^\mathbf{R} + \boldsymbol{\sigma}^\mathbf{R} \right)_{rs}.
\end{align}
Note that we are not using the symmetrized transition density matrix here.

The formalism for the CI term of the SA-CASSCF derivative coupling is similar to that of the SA-CASSCF nuclear gradient.
In other words, the extension of the nuclear gradient algorithm to derivative coupling is straightforward in the SA-CASSCF case.
The MRCI derivative coupling\cite{Lengsfield1984JCP,Lengsfield1992ACP}
and semiempirical derivative coupling\cite{Toniolo2002JPCA,Toniolo2003JPCA,Fabiano2008CP} can be evaluated without significant modifications in the formalism as well.
The use of the density-fitting scheme \cite{Galvan2016JCTC}
and the GPU algorithms for SA-CASSCF derivative coupling have recently been reported.\cite{Snyder2015JCP,Snyder2017JCP}

\section{Multireference Electron Correlation Gradients}\label{sec_mr_gradient}

Now, we review analytical gradient theory for internally contracted multireference electron correlation methods.
Direct differentiation is often not straightforward,
since there are many terms and the contractions vary with respect to the nuclear positions.
As an alternative, the Lagrangian approach is used.\cite{Pulay1983JCP,King1986JCP,Pulay1995bookchapter,Koch1990JCP2,Jorgensen1988JCP,Celani2003JCP}
First, we introduce the Lagrangian in the single-reference case.\cite{Pulay1983JCP,King1986JCP,Pulay1995bookchapter,Koch1990JCP2,Jorgensen1988JCP}
We shall see that the Lagrangian approach yields the same gradient as in direct differentiation.
The multireference extension of the Lagrangian approach is introduced next.
Finally, the formulation of the nuclear gradients and properties for the multireference electron correlation models
are reviewed
with various internal contraction schemes and selection of the zeroth-order Hamiltonians.

\subsection{Direct differentiation}

It is possible to use direct differentiation to derive the analytical gradient formalism particularly for uncontracted MRCI methods.
In fact, the MRCI energy is computed by minimizing the CI energy without optimizing the orbitals,
and it is very natural that the gradient can be evaluated in a way that is similar to the SA-CASSCF nuclear gradients.\cite{Page1984JCP,Rice1985CPL}
A program for computing the MRCI nuclear gradients, based on these formulae, was developed by Shepard and coworkers,\cite{Shepard1987IJQC,Shepard1992JCP}
and was later extended to the size-consistency corrected MR-AQCC method\cite{Lischka2001PCCP} and
those with state-averaged reference functions by Lischka {\it{et al.}}\cite{Lischka2002MP}
The derivative coupling between the MRCI states has been evaluated employing a similar approach.\cite{Lengsfield1984JCP,Saxe1985CPL,Lischka2004JCP}
There is a work that reports evaluation the MRCI analytical gradient using the response function, or the Lagrangian approach.\cite{Khait2010MP}
Very recently, the analytical gradient of one of the variants of NEVPT2 using direct differentiation has been reported,\cite{Nishimoto2019Chemrxiv} 
although the same working equations have been derived using the Lagrangian approach as well.\cite{Park2019ArXiv}

\subsection{The Lagrangian approach: The Single-Reference Case}

In the Lagrangian approach, the optimization criteria are used as constraints.\cite{Pulay1983JCP,King1986JCP,Pulay1995bookchapter,Koch1990JCP2,Jorgensen1988JCP}
As the simplest example, the Hartree--Fock Lagrangian is
\begin{align}
\mathcal{L} = \sum_{\mu\nu} d_{\mu\nu} h_{\mu\nu} + \frac{1}{2} \sum_{\mu\nu\rho\lambda} D_{\mu\nu,\rho\lambda} (\mu\nu|\rho\lambda) - \frac{1}{2}\tr\left[\mathbf{X}\left(\mathbf{C}^\dagger \mathbf{SC} - \mathbf{1}\right)\right],
\end{align}
where the expression $\mathbf{C}^\dagger \mathbf{SC} = \mathbf{1}$ is associated with the orthonormality of the molecular orbitals;
we have the unknown multiplier $\mathbf{X}$ for this condition.
The multiplier $\mathbf{X}$ is to be obtained by satisfying the stationary condition of $\mathcal{L}$ with respect to the variation of $\mathbf{U}$:
\begin{align}
\frac{\partial{\mathcal{L}}}{\partial U_{xy}} = 2 \left[ \sum_{j} d_{yj} h_{xj} + \sum_{jkl} D_{yj,kl} (xj|kl)  \right] - X_{xy} = 0,
\end{align}
which yields $\mathbf{X}$ to be
\begin{align}
X_{xy} = 
\left\{ \begin{array}{ll}
\displaystyle
4 \epsilon_{x} \delta_{xy}
& {x,y} \in \mathrm{occupied} \\[8pt]
\displaystyle
0
& \mathrm{otherwise}
\end{array} \right.
\end{align}
in the closed-shell case.
The nuclear gradient is evaluated using the Lagrangian with the molecular integrals replaced by their derivative integrals, as
\begin{align}
\frac{\mathrm{d} E}{\mathrm{d} \mathbf{R}}
& = \frac{\partial \mathcal{L}}{\partial \mathbf{R}} \nonumber \\
& = \sum_{\mu\nu} d_{\mu\nu} h_{\mu\nu}^\mathbf{R}
+ \frac{1}{2} \sum_{\mu\nu\rho\lambda} D_{\mu\nu,\rho\lambda} (\mu\nu|\rho\lambda)^\mathbf{R} -
\frac{1}{2}\tr\left[\mathbf{X}\mathbf{C}^\dagger \mathbf{S^{R}C}\right].
\end{align}
Inserting the explicit expression for $\mathbf{X}$ gives
an expression equivalent to Eq.~\eqref{hfderiv}.

The Lagrangian approach naturally leads to the Z-vector algorithms.
For example,
the Lagrangian for MP2 is\cite{Jorgensen1988JCP}
\begin{align}
\mathcal{L} = E^\mathrm{MP2} - \frac{1}{2} \sum_{i}^{\mathrm{occ}}\sum_{a}^{\mathrm{vir}} Z_{ai} f_{ai} - \frac{1}{2} \tr \left[ \mathbf{X} \left( \mathbf{C}^\dagger \mathbf{SC} - \mathbf{1} \right) \right],
\end{align}
where $f_{ai} = 0$ defines the Brillouin condition.
$Z$ is the Lagrangian multiplier for this constraint.
Differentiating the Lagrangian with respect to $U_{rs}$ yields
\begin{align}
\frac{\partial \mathcal{L}}{\partial U_{rs}} - \frac{\partial \mathcal{L}}{\partial U_{sr}} & = Y_{rs} - Y_{sr} - \frac{1}{2} \sum_{i}^{\mathrm{occ}}\sum_{a}^{\mathrm{vir}} Z_{ai} \left( \frac{\partial f_{ai}}{\partial U_{rs}}  - \frac{\partial f_{ai}}{\partial U_{sr}}\right) \nonumber \\
& = Y_{rs} - Y_{sr} - Z_{ai} A_{airs}= 0,
\end{align}
where $Y_{rs} = \partial E_\mathrm{MP2} / \partial U_{rs}$.
Note that the explicit antisymmetrization removes the contributions from the last term.
This is equivalent to the $Z$-vector equation, Eq.~\eqref{srzvector}.
$\mathbf{X}$ is obtained from the symmetric part as
\begin{align}
X_{rs} = Y_{rs} - \frac{1}{2}\sum_{i}^{\mathrm{occ}}\sum_{a}^{\mathrm{vir}} Z_{ai} A_{airs}.
\end{align}
The second term in the right-hand side accounts for the overlap-related terms in $\mathbf{B}^{\mathbf{R}}$ in Eq.~\eqref{handyshaefer},
and the resulting nuclear gradient formulae remain identical to Eq.~\eqref{correlderiv}.

\subsection{The Lagrangian Approach: The Multireference Case}

One of the most elegant demonstrations of the Lagrangian for multireference nuclear gradients was the work by Celani and Werner
for partially internally contracted CASPT2 nuclear gradients.\cite{Celani2003JCP}
The Celani--Werner Lagrangian approach is applicable for all multireference methods
and is mathematically equivalent to the other forms of the Lagrangian for other (uncontracted) multireference theories such as MRCI, MRCC, MCQDPT2, GVVPT2, or NEVPT2,
which will be introduced in Sec.~\ref{sec_otherforms}.
Here, we will first introduce the Lagrangian formalism for CASSCF theory,
which yields essentially the same results as the direct differentiation in the previous section.

The Lagrangian for the single-state CASSCF gradient has a form similar to the Hartree--Fock one, as
\begin{align}
\mathcal{L} & = \sum_{\mu\nu} d_{\mu\nu} h_{\mu\nu} + \frac{1}{2} \sum_{\mu\nu\rho\lambda} D_{\mu\nu,\rho\lambda} (\mu\nu|\rho\lambda) \nonumber \\ & - \frac{1}{2}\tr\left[\mathbf{X}\left(\mathbf{C}^\dagger \mathbf{SC} - \mathbf{1}\right)\right] - \frac{1}{2} x \left( \langle \Psi_0 | \Psi_0 \rangle - 1 \right).
\end{align}
The only additional constraint is the orthonormality of the reference function.
The multipliers are determined to be
\begin{align}
& X_{ij} = 2 \left( h_{ij} d_{ij} + \sum_{kl}D_{ij,kl} (ij|kl) \right), \\
& x = 0.
\end{align}
The multipliers (and therefore, nuclear gradients) reduce to the Hartree--Fock gradient when there are no active orbitals.

Because individual state energies in SA-CASSCF are not stationary with respect to the orbital and CI coefficients,
the Lagrangian for SA-CASSCF now has the conditions for CASSCF convergence,
\begin{align}
& \mathbf{A} - \mathbf{A}^\dagger = \mathbf{0}, \\
& \hat{H} | N \rangle = E_N | N \rangle,
\end{align}
where
\begin{align}
A_{rs} & = 
\left\{ \begin{array}{ll}
\displaystyle
2 \sum_{j} h_{rj} d_{sj}^\mathrm{SA} + 2 \sum_{jkl} J_{rj}^{kl} D_{sj,kl}^\mathrm{SA}
& {s} \in \mathrm{occupied} \\[8pt]
\displaystyle
0
& \mathrm{otherwise}
\end{array} \right.
\end{align}
and the superscript $\mathrm{SA}$ means that the density matrices are state-averaged, i.e.,
\begin{align}
& d^\mathrm{SA}_{rs} = \sum_N W_N \langle N | \hat{E}_{rs} | N \rangle, \\
& D^\mathrm{SA}_{rs,tu} = \sum_N W_N \langle N | \hat{E}_{rs,tu} | N \rangle.
\end{align}
Incorporating these conditions into the Lagrangian leads to the expression
\begin{align}
\mathcal{L} & = E_\mathrm{SA-CASSCF} \nonumber \\
& + \frac{1}{2}\tr\left[\mathbf{Z}^\dagger (\mathbf{A}-\mathbf{A}^\dagger)\right] - \frac{1}{2} \tr \left[ \mathbf{X} \left( \mathbf{C}^\dagger \mathbf{SC} - \mathbf{1} \right) \right] \nonumber \\
& + \sum_N W_N \left[ \sum_I z_{I,N} \langle I | \hat{H}-E_N | N \rangle - \frac{1}{2} x_N \left( \langle N | N \rangle - 1 \right) \right].
\end{align}
The corresponding multipliers are obtained by solving the $Z$-vector equations, which are
\begin{align}
& \frac{\partial\mathcal{L}}{\partial U_{rs}} - \frac{\partial\mathcal{L}}{\partial U_{rs}}
= \left( 1 - \tau_{rs} \right) \left[ \mathbf{Y} + \bar{\bar{\mathbf{A}}} + \tilde{\mathbf{A}} \right]_{rs} = 0 \label{Zeq}, \\
& \frac{1}{W_N }\frac{\partial\mathcal{L}}{\partial c_{I,N}} = \frac{y_{I,N}}{W_N} + \left(\mathbf{H} - E_N\right) \mathbf{z}^N  - x_N \mathbf{c}^N + 2 \tilde{\mathbf{H}} \mathbf{c}^N = 0, \label{zeq}
\end{align}
where $\tau_{rs}$ permutes the indices $r$ and $s$,
and $\bar{\bar{A}}$ and $\tilde{A}$ are $\mathbf{A}$ matrix evaluated with
the $z$-weighted symmetrized density matrix and the $Z$-weighted molecular integrals, respectively,
\begin{align}
& \bar{\bar{A}}_{rs} = 2 \sum_{j} h_{rj} \bar{\bar{d}}_{sj}^\mathrm{SA} + 2 \sum_{jkl} J_{rj}^{kl} \bar{\bar{D}}_{sj,kl}^\mathrm{SA} & {s} \in \mathrm{occupied}\\
& \tilde{A}_{rs} = 2 \sum_{j} \tilde{h}_{rj} d_{sj}^\mathrm{SA} + 2 \sum_{jkl} \tilde{J}_{rj}^{kl} D_{sj,kl}^\mathrm{SA} & {s} \in \mathrm{occupied}
\end{align}
and
\begin{align}
& \bar{\bar{d}}_{rs}^\mathrm{SA} = \frac{1}{2} \sum_N W_N \sum_I z_{I,N} \langle I | \hat{E}_{rs} + \hat{E}_{sr} | N \rangle, \\
& \bar{\bar{D}}_{rs,tu}^\mathrm{SA} = \frac{1}{2} \sum_N W_N \sum_I z_{I,N} \langle I | \hat{E}_{rs,tu} + \hat{E}_{tu,sr} | N \rangle, \\
& \tilde{h}_{rj} = \sum_{s} \left[ Z_{rs} h_{sj} + h_{rs} Z_{sj} \right] \\
& \tilde{J}_{rj}^{kl} = \sum_{s} \left[ Z_{sr} (sj|kl) + Z_{sj} (rs|kl) + Z_{sk} (rj|sl) + Z_{sl} (rj|ks) \right].
\end{align}
This equation is equivalent to the $Z$-vector equation obtained by direct differentiation, Eq.~\eqref{zcasscf_direct}.
The source terms for the $Z$-vector equations,
which are the derivative of the energy with respect to the orbital rotation coefficients and the CI coefficient,
should be calculated.
The orbital derivative $Y$ is in Eq.~\eqref{orbderivsacasscf},
while the derivative with respect to the CI coefficient (CI derivative) $y_{I,N}$ is zero. 
The coupled equations Eq.~\eqref{Zeq} and \eqref{zeq} are iteratively solved to find $\mathbf{Z}$, $\mathbf{z}$ and $\mathbf{X}$.
Using these multipliers, the effective densities are computed,
and the gradient is evaluated by Eq.~\eqref{effectivedensity}.

\subsection{CASPT2 Analytical Gradient and the Automatic Implementation Technique}

The Celani--Werner Lagrangian was originally suggested for the CASPT2 gradient.\cite{Celani2003JCP}
It reads
\begin{align}
\mathcal{L} & = E_{\mathrm{CASPT2}} \nonumber \\
& + \frac{1}{2}\tr\left[\mathbf{Z}^\dagger (\mathbf{A}-\mathbf{A}^\dagger)\right] - \frac{1}{2} \tr \left[ \mathbf{X} \left( \mathbf{C}^\dagger \mathbf{SC} - \mathbf{1} \right) \right] \nonumber \\
& + \sum_N W_N \left[ \sum_I z_{I,N} \langle I | \hat{H}-E_N | N \rangle - \frac{1}{2} x_N \left( \langle N | N \rangle - 1 \right) \right] \nonumber \\
& + \sum_i^{\mathrm{inactive}} \sum_{j}^{\mathrm{core}} z_{ij} f_{ij},\label{cwlagrangian}
\end{align}
where the last term accounts for the frozen core approximation.\cite{Celani2003JCP}
The Lagrangian was written for the partially contracted variant of CASPT2 (WK-CASPT2, see Sec.~\ref{sec_caspt2}), but 
is equally applicable to the internally contracted CASPT2 as well.
The same $Z$-vector equation should be solved with the different source terms.
As mentioned in Sec.~\ref{sec_caspt2},
the CASPT2 energy results from the minimization of the Hylleraas functional,
\begin{align}
E^\mathrm{CASPT2}
&= E^\mathrm{CASSCF} \nonumber \\
&+ \langle \Phi^{(1)} | \hat{H}^{(0)} - E^{(0)} | \Phi^{(1)} \rangle + \langle \Phi^{(1)} | \hat{H} | \Phi_0 \rangle.
\end{align}
The energy derivative with respect to the orbital rotation coefficients
is computed in a straightforward manner by recasting the energy as
\begin{align}
E^\mathrm{CASPT2} = \tr \left(\mathbf{hd}\right) + \tr\left[\mathbf{g}(\mathbf{d}^{(0)})\mathbf{d}^{(2)}\right] +
\tr \sum_{kl} \left( \mathbf{K}^{kl} \mathbf{D}^{lk}\right),
\end{align}
where the shorthand notations for the integrals,
\begin{align}
\left[\mathbf{g}(\mathbf{d})\right]_{xy} & = \sum_{kl} \left[ (xy|zw) d_{zw} - \frac{1}{4}(xw|zy) (d_{zw} + d_{wz}) \right],\\
\mathbf{K}^{zw}_{xy} & = (xz|yw),
\end{align}
were employed.
The one-electron and two-electron density matrices are
summations of the zeroth, first, and second-order density matrices,
\begin{align}
& \mathbf{d} = \mathbf{d}^{(0)} + \mathbf{d}^{(1)} + \mathbf{d}^{(2)}, \\
& \mathbf{D} = \mathbf{D}^{(0)} + \mathbf{D}^{(1)},
\end{align}
where the perturbation orders of the density matrices are
the number of amplitude terms in the density matrices.
The derivative of the energy with respect to the energy coefficient is
\begin{align}
\frac{1}{2} Y_{ri} & = \left[\mathbf{hd} + \mathbf{g}(\mathbf{d}^{(0)}) \mathbf{d}^{(2)} + \mathbf{g}(\mathbf{d}^{(2)}) \mathbf{d}^{(0)} + \sum_{kl} \mathbf{K}^{kl} \mathbf{D}^{lk} \right] \nonumber \\& + \sum_{jst} D_{st}^{ij} (rs | tj), \\
\frac{1}{2}Y_{ra} & = \left[\mathbf{hd} + \mathbf{g}(\mathbf{d}^{(0)}) \mathbf{d}^{(2)} + \sum_{kl} \mathbf{K}^{kl} \mathbf{D}^{lk} \right]_{ra},
\end{align}
The explicit form of the density matrices and the derivatives with respect to the CI coefficient depends on the internal contraction scheme.

The first application of the Celani--Werner Lagrangian toward gradient theory\cite{Celani2003JCP} was for partially internally contracted CASPT2 (PIC-CASPT2).
Although the explicit expressions of the density matrices and CI derivatives are quite involved,
they are derived by hand and the explicit expressions are shown in the literature.\cite{Celani2003JCP,Shiozaki2011JCP3}
The density-fitting scheme is used for evaluating the source terms.\cite{Gyorffy2013JCP}

On the other hand, 
in fully internally contracted CASPT2,
all the two-electron excitations from the reference functions
are contracted.
This allows for a smaller number of the amplitudes,
and a simplified mathematical formula by using the operator $\hat{T}$.
However, the internal contraction of singly external configurations
requires 8-index quantities in the active space.
For FIC-CASPT2, the CI derivative is
\begin{align}
\frac{1}{2}y_I & = \langle I | \hat{H} | \Phi_0 \rangle + \langle I | \hat{T}^\dagger (\hat{H}^{(0)} - E^{(0)}) \hat{T} | \Phi_0 \rangle \nonumber \\
& + \langle I | \hat{T}^\dagger \hat{H} | \Phi_0 \rangle
+ \sum_{rs} \langle I | \hat{E}_{rs} | \Phi_0 \rangle \left[ \mathbf{g}(\mathbf{d}^{(2)}) - N \mathbf{f} \right]_{rs},
\end{align}
where the density matrices are
\begin{align}
&d^{(1)}_{xy} = \langle \Phi_0 | \hat{T}^\dagger \hat{E}_{xy} | \Phi_0 \rangle, \\
&d^{(2)}_{xy} = 
\left\{ \begin{array}{ll}
\displaystyle
\langle \Phi_0 | \hat{T}^\dagger \hat{E}_{xy} \hat{T}| \Phi_0 \rangle - N \langle \tilde{L} | \hat{E}_{xy} | \tilde{L} \rangle
& {x,y} \in {r,s} \\[8pt]
\displaystyle
\langle \Phi_0 | \hat{T}^\dagger \hat{E}_{xy} \hat{T}| \Phi_0 \rangle
& \mathrm{otherwise}
\end{array} \right. \\
&D^{(1)}_{xy,zw} = \langle \Phi_0 | \hat{T}^\dagger \hat{E}_{xy,zw} | \Phi_0 \rangle,
\end{align}
and the correlated norm is defined as
\begin{align}
N = \langle \Psi_0 | \hat{T}^\dagger \hat{T} | \Psi_0 \rangle.
\end{align}
The tensor operations in the FIC-CASPT2 nuclear gradient are quite involved,
and manual implementations of the second-order density matrix and the CI derivatives had been an obstacle.

This technical challenge was addressed by MacLeod and Shiozaki in 2015 by using the automatic implementation technique.\cite{MacLeod2015JCP}
This technique is dominantly used to implement higher-order CC, MRCI, and complicated nuclear gradient algorithms for effective core potentials (ECP).\cite{Hirata2006TCA,Kallay2001JCP,Hirata2003JPCA,Hirata2004JCP,Hanrath2010JCP,Engels-Putzka2011JCP,Shiozaki2008PCCP,Shiozaki2008JCP,Kohn2008JCP,Shiozaki2009JCP,Nataraj2010JCP,Rolik2011JCP,Kats2013JCP,Datta2013JCTC,Datta2014JCP,Wladyslawski2005AQC,Neuscamman2009JCP,Parkhill2009JCP,Parkhill2010MP,Hanauer2011JCP,Saitow2013JCP,SongJCTC2016}
The automatic code generator program {\sc smith3} exploits Wick's theorem to rearrange the operators
to arrive at the expressions with simple tensor contractions and RDMs in the active space.
The code generator treats the determinant index $I$ in an analogous way to the orbitals,\cite{MacLeod2015JCP}
and makes it easy to evaluate the CI derivatives.

The original algorithm written by {\sc smith3} was later modified to accelerate the computations of the CI derivative.\cite{Park2017JCTC2}
The CI derivative was recast into
\begin{align}
y_I & = A^{(0)} \Gamma^{(0)I} + \sum_{rs} A^{(1)}_{rs} \Gamma^{(1)I}_{rs}
+ \sum_{rstu} A^{(2)}_{rs,tu} \Gamma^{(2)I}_{rs,tu} \nonumber \\
& + \sum_{rstuvw} A^{(3)}_{rs,tu,vw} \Gamma^{(3)I}_{rs,tu,vw}
+ \sum_{rstuvw} A^{(4)f}_{rs,tu,vw} \Gamma^{(4)fI}_{rs,tu,vw},\label{yi_rdmrecast}
\end{align}
where $\Gamma^{(n)I}$ denotes the derivative of the $n$-th order RDM with respect to the CI coefficient (which are RDM derivatives), e.g.,
\begin{align}
& \Gamma^{(3)I}_{rs,tu,vw} = \langle I | \hat{E}_{rs,tu,vw} | \Phi_0 \rangle, \nonumber \\
& \Gamma^{(4)fI}_{rs,tu,vw} = \sum_{xy} \langle I | \hat{E}_{rs,tu,vw,xy} | \Phi_0 \rangle f_{xy},
\end{align}
where $f_{xy}$ is an element of the Fock matrix.
The RDM derivatives require the evaluation of the last line of Eq.~\eqref{yi_rdmrecast} and are evaluated using the direct algorithm,
which is similar to the Knowles--Handy algorithm for determinant-based FCI.\cite{Knowles1984CPL}
This results in a reduced memory requirement
and improved computational efficiency.

\subsection{Multistate CASPT2 Gradient and Derivative Coupling}

Next, let us discuss the multistate extension of CASPT2 gradient theory.
The (X)MS-CASPT2 energies are not stationary with respect to the amplitudes.\cite{Shiozaki2011JCP3,Vlaisavljevich2016JCTC}
The Lagrangian, which is stationary with respect to the changes of the amplitude, is written as
\begin{align}
\mathcal{L}^\mathrm{PT2}_Q & = E_Q^{\mathrm{MS-CASPT2}} \nonumber \\ &+ 
\sum_N \sum_{\Omega} \lambda_{\Omega,N} \left[ \langle \Omega | \hat{H}^{(0)} - E_N^{(0)} | N \rangle + \langle \Omega | \hat{H} | N \rangle \right].\label{lambda}
\end{align}
The suitable values of $\lambda_{\Omega, N}$ are to be determined by the so-called $\lambda$-equation,
making this Lagrangian stationary with respect to the variation of both $T_{\Omega, N}$ and $\lambda_{\Omega, N}$.
The $\lambda$-equation is obtained by differentiating the Lagrangian
with respect to the amplitudes.
The full Celani--Werner Lagrangian is essentially the same as
that for the single-state case [Eq.~\eqref{cwlagrangian}],
except for the fact that the energy is substituted by the PT2 Lagrangian $\mathcal{L}^\mathrm{PT2}$.
The density matrices and CI derivatives,
which are the source terms for the $Z$-vector equation,
are also accordingly modified to reflect these changes.
They are derived by hand for PIC-CASPT2\cite{Shiozaki2011JCP3} or by automatic implementation technique for FIC-CASPT2.\cite{Vlaisavljevich2016JCTC}

It is also possible to evaluate the derivative coupling with (X)MS-CASPT2 using the same program infrastructure,
because the final (X)MS-CASPT2 energy is obtained by the diagonalization of the effective Hamiltonian. 
The interstate coupling, which is part of the derivative coupling, is
\begin{align}
\mathbf{h}^{QP}_{\mathrm{interstate}} = R_{MQ} \frac{\mathrm{d} H^\mathrm{eff}_{MN}}{\mathrm{d} \mathbf{X}} R_{NP},
\end{align}
and was implemented for PIC-MS-CASPT2 to find the conical intersection of cyclohexadiene and PSB3 at the MS-CASPT2 level,\cite{Mori2009CPL,Mori2010JCP}
using the PIC-MS-CASPT2 nuclear gradient program developed in {\sc molpro}.
The Lagrangian [Eq.~\eqref{lambda}] is modified 
as
\begin{align}
\mathcal{L}^\mathrm{PT2}_{QP} & = R_{MQ }H_{MN}^{\mathrm{eff}} R_{NP} \nonumber \\ &+ 
\sum_N \sum_{\Omega} \lambda_{\Omega,N} \left[ \langle \Omega | \hat{H}^{(0)} - E_N^{(0)} | N \rangle + \langle \Omega | \hat{H} | N \rangle \right].\label{lambda_dcv}
\end{align}
From the perspective of the time-independent Schr\"odinger equation, however,
the derivative coupling between the multistate CASPT2 states $P$ and $Q$ is defined as
\begin{align}
\mathbf{h}^{\mathrm{PT2,}QP} = \frac{1}{2}
\left[ \braket{Q}{\frac{\mathrm{d}\Psi_P}{\mathrm{d}\mathbf{X}}}
+ \braket{\Psi_Q}{\frac{\mathrm{d}P}{\mathrm{d}\mathbf{X}}}\right],
\end{align}
and requires two more terms that account for the derivatives of the XMS rotation term and the first-order correction to the wave function.
These terms are evaluated by simple modifications of the $Z$-vector equation.\cite{Park2017JCTC}

\subsection{Shifts and selection of Fock operators}

As we have mentioned in Sec.~\ref{sec_caspt2},
there are various definitions for the zeroth-order Hamiltonian in the CASPT2 method.
The real and imaginary level shift were both introduced to alleviate the intruder state problem.
The imaginary shift uses the real part of the complex amplitudes to avoid the use of complex algebra in energy evaluation.\cite{Forsberg1997CPL}
Its analytical nuclear gradients have been recently reported.\cite{Park2019JCTC2}
It was found that, compared to the real shift, the imaginary shift improves the accuracy at the cost of additional computational burden of evaluating effective density matrix elements.

\begin{figure}[tb]
	\includegraphics[width=0.9\linewidth]{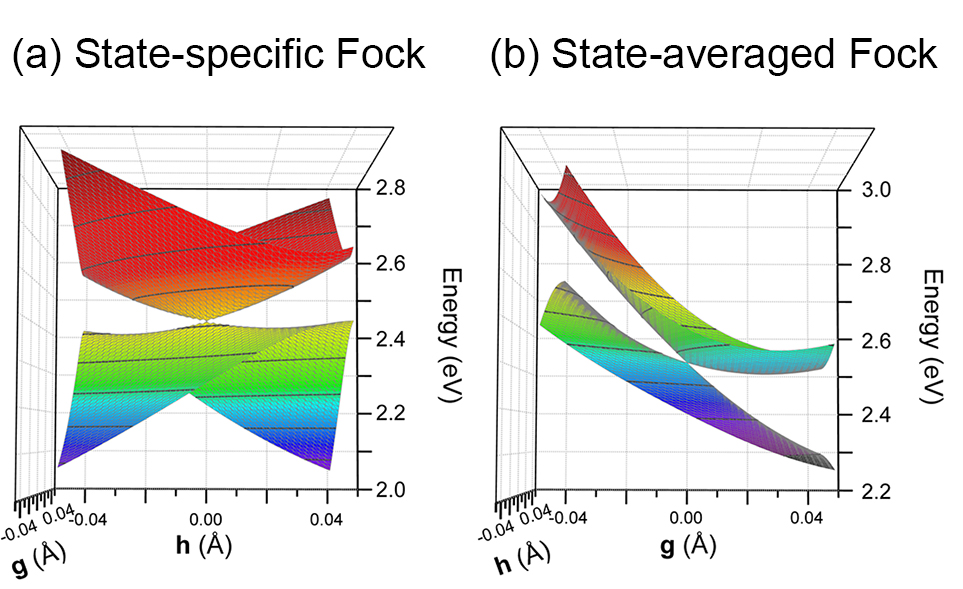}
	\caption{PES contour plots near the MECIs of PSB3 computed with different methods. The MECIs obtained with state-specific Fock operator and state-averaged Fock operator have peaked and sloped topologies, respectively. \label{figure:sssrmsmr}}
\end{figure}

There are two types of Fock operators from which zeroth-order Hamiltonians for MS-CASPT2 in Eqs.~\eqref{sssrmscaspt2} and \eqref{msmrmscaspt2} are constructed: state-speficic
and state-averaged Fock operators.
The analytical nuclear gradient program for CASPT2 with the state-specific Fock operator was recently developed to test its accuracy.\cite{Park2019JCTC}
Notably the geometries and the topologies of the MECIs qualitatively change with the choice of the Fock operator in the definition of zeroth-order Hamiltonian,
even if both use SS-SR internal contraction (Fig.~\ref{figure:sssrmsmr}). 
The difference between these zeroth-order Hamiltonians was also demonstrated by Mart{\'i}nez and coworkers.\cite{Liu2019ChemRxiv}

\subsection{Uses of the Lagrangian in Other Multireference Models}\label{sec_otherforms}

Similar Lagrangians have since been used for other multireference theories,
even though notations and conventions are often different from the original work by Celani and Werner.
For example, the use of the Lagrangian approach by Helgaker\cite{Helgaker1989TCA} for molecular properties with single-state MRCISD predates the work by Celani and Werner. 
The multistate MRCISD nuclear gradient and derivative coupling were formulated using a similar Lagrangian.\cite{Khait2010MP}
In the context of perturbation theories,
the MCQDPT2 gradient,\cite{Nakano1998JCP} the GVVPT2 gradient,\cite{Theis2011JCP} and derivative coupling,\cite{Khait2012CP}
and the single-state NEVPT2 gradient\cite{Park2019ArXiv}
were formulated using the same Lagrangian.
We note that neither MCQDPT2 nor GVVPT2 uses internal contraction, making the equations amenable to manual derivation.
For the MRCC methods, the analytical gradient methods for two-determinant CCSD,\cite{Szalay1994JCP}
the Brillouin--Wigner MRCC method,\cite{Pittner2007JCP}
and the Mukherjee MRCC methods\cite{Prochnow2009JCP,Jagau2010JCP}
have been reported.
 
\section{Interface to Photodynamics/ Photochemistry}\label{sec_interface}

Using the methodologies discussed in the previous section,
it is possible to evaluate the analytical gradient and derivative coupling within the framework of multireference electronic structure theories.
However, without applications in real chemical systems discussed in Secs.~\ref{sec_application_geomopt} and \ref{sec_application_dynamics},
the importance of nuclear gradient theory will be not so significant.
In this section,
we will review how the analytical gradient and derivative couplings are used in such applications for optimizing conical intersections and performing nonadiabatic 
dynamics
simulations.
Detailed reviews on this topic can be found in Refs.\onlinecite{Baerbook,CIbook,CrespoOtero2018CR}.

\subsection{Derivative Coupling and Conical Intersection Optimization}

Conical intersections are points where two different electronic states are degenerate in energy, or where potential energy surfaces meet.
They are involved in photochemistry/ photodynamics, where different electronic states are involved,
where phenomena beyond the Born--Oppenheimer approximation usually occur.\cite{Baerbook, CIbook}
Multireference methods can be used to optimize the geometries of conical intersections.
When using the term optimization for conical intersections,
we should be careful,
because conical intersections are multidimensional on the PES
unlike zero-dimensional equilibrium geometries (the point where the potential energy is at a minimum)
or transition state geometries (the saddle point on the PES).
This fact implies that, apart from the condition that two (or more)\cite{Katriel1980CPL,Matsika2005JPCA} electronic states should be isoenergetic, one should define an additional constraint,
so that the optimization procedure will look for only one geometry.

The most widely used condition is the minimal energy condition.
The resulting geometrical point is the minimal energy conical intersection (MECI),
which is the minimum energy point along the conical intersection seam.
When there are two states crossing,
the
MECI optimization is a minimization of the energy of the upper state until the energy difference between the two states is zero.\cite{Keal2007TCA}
Naturally, there are two vectors that lift degeneracy between the two electronic states $P$ and $Q$:
The gradient difference vector,
\begin{align}
\mathbf{g}^{PQ} = \frac{\mathrm{d}}{\mathrm{d}\mathbf{R}} \left( \langle \Psi_P | \hat{H} | \Psi_P \rangle - \langle \Psi_Q | \hat{H} | \Psi_Q \rangle \right),
\end{align}
and the interstate coupling vector,
\begin{align}
\mathbf{h}^{PQ} = \left\langle \Psi_P \left| \frac{\mathrm{d}\hat{H}}{\mathrm{d}\mathbf{R}} \right| \Psi_Q \right\rangle.
\end{align}
For multireference electron correlation methods,
these vectors can be obtained by the methods described in the previous section.
We note that the interstate coupling vector gives a major contribution to the derivative coupling vector near the conical intersection, but they are not necessarily the same.
These two vectors play a crucial role in optimizing MECIs.\cite{Keal2007TCA}
Since many quantum chemical methods only have the algorithm for state-specific nuclear gradients
(which means, only knowledge of $\mathbf{g}^{PQ}$ is available),
there are algorithms that do not require $\mathbf{h}^{PQ}$,
which optimizes the functions defined only with the averaged energy, energy differences, and some empirical parameters.\cite{Ciminelli2004ChemEurJ,Levine2006MP,Levine2008JPCB}
When $\mathbf{h}^{PQ}$ is available, the efficiency of the optimization is substantially improved.\cite{Keal2007TCA,Levine2008JPCB}
These algorithms include the Lagrange--Newton type constrained optimization methods,\cite{Koga1985CPL,Farazdel1991JCC,Manaa1993JCP,Yarkony1993JPC,Anglada1997JCC}
and the gradient projection method by Bearpark \textit{et al.}\cite{Bearpark1994CPL}
These algorithms are being continuously improved over time.\cite{Toniolo2002JPCA,Yarkony2004JPCA,Sicilia2008JCTC,RuizBarragan2013JCTC,DeVico2005JCTC}

In 2008, Mart{\'i}nez and coworkers suggested the concept of the minimal distance conical intersection (MDCI).\cite{Levine2008JPCB}
In the MDCI search, one optimizes the conical intersection point that is the closest to the reference geometry,
such as the excited-state minimum or the Franck--Condon point.
The MDCI optimization is done with and without $\mathbf{h}^{PQ}$
as in the case of the MECI optimization.\cite{Levine2008JPCB}

The conical intersection is a seam rather than a zero-dimensional point,
and the structures of the seam can be very complicated.
Naturally,
it benefits greatly from an automated search procedure.
Recently, such automated search algorithms were implemented for finding minimal energy points in the seam,\cite{Maeda2009JPCA,Maeda2011JPCL,Maeda2012AdvPhysChem,Harabuchi2013JCTC,Maeda2014JPCA,Maeda2015JACS,Mori2013JCTC,Aldaz2018PCCP,Lindner2018PRA,Lindner2019JCTC}
which may be regarded as extensions of automated transition state structure search algorithms\cite{Maeda2009JPCA,Maeda2011JPCL,Maeda2012AdvPhysChem,Harabuchi2013JCTC,Maeda2014JPCA,Maeda2015JACS,Mori2013JCTC,Aldaz2018PCCP}
or metadynamics.\cite{Lindner2018PRA,Lindner2019JCTC}
This is one of the areas of active development.

\subsection{Semiclassical Dynamics Simulations}

When possible, nonadiabatic (semiclassical or mixed quantum-classical)
dynamics simulations yield a great deal of information and intuition for photochemical/ photodynamic processes.
There have been a myriad of nonadiabatic dynamics simulation methods developed over the last 40 years.\cite{CrespoOtero2018CR,Curchod2018CR,Tully1990JCP,Hammes-Schiffer1994JCP,Martens1997JCP,Kapral1999JCP,MacKernan2008JPCB,Stock1997PRL,Thoss1999PRA,Ben-Nun2000JPCA,Kelly2012JCP,Tavernelli2015ACR,Tavernelli2013PRA,Curchod2011PCCP,Agostini2016JCTC,Barbatti2011WIREs,Subotnik2016ARPC,Yonehara2012CR,Virshup2009JPCB}
In all molecular dynamics (trajectory) simulations, no matter whether there are nonadiabatic transitions or not,
the nuclei are treated in a classical-mechanical way.
This means that
Newton's equation of motion,
\begin{align}
\mathbf{F} = m \mathbf{a} = -\nabla V = \frac{\mathrm{d}}{\mathrm{d}\mathbf{X}} \langle \Psi | \hat{H}| \Psi \rangle,
\end{align}
are numerically integrated for each nucleus at each time step,
and the nuclear gradient is again needed in propagation of the trajectories.
To include a nonadiabatic transition,
the coupling between the nuclear motion and electronic motion,
which breaks the Born--Oppenheimer approximation,
should be taken into account with the derivative coupling.
For example, in the FSSH algorithm, \cite{Tully1990JCP,Hammes-Schiffer1994JCP,Barbatti2011WIREs,Yonehara2012CR,Subotnik2016ARPC} 
the quantum mechanical amplitude of the electronic state $P$, $c_P$, is given by
\begin{align}
i \hbar \frac{\mathrm{d} c_P(t)}{\mathrm{d} t} =
c_P(t) E_P - i \hbar \sum_Q c_Q(t) \frac{\mathrm{d} \mathbf{X}}{\mathrm{d} t} \left\langle \Psi_P \left| \frac{\mathrm{d}}{\mathrm{d}\mathbf{X}} \right| \Psi_Q\right\rangle.
\end{align}
It is easy to see that the second term includes the nuclear motion ($\mathrm{d}\mathbf{X}/\mathrm{d}t$)
and
that the
electronic motion
(the other electronic states given by $c_Q$) is coupled by the derivative coupling.
In all semiclassical methods,
though the mathematical way of treating the nonadiabatic transitions is different,
computations of the potential energy, energy gradient, and derivative coupling are needed at each time step,
as long as the nuclear motion is classically treated.\cite{Yonehara2012CR,CrespoOtero2018CR,Curchod2018CR}
Therefore, semiclassical dynamics simulations are usually conducted in a following way:
\begin{enumerate}
	\item The initial geometry and velocity for the trajectory is given.
	\item Evaluate the energy, nuclear gradient, and derivative coupling and pass them to the trajectory propagator.\label{evaluate_energy}
	\item The trajectory, including the nuclear position and quantum mechanical density for the electronic states, is propagated.\label{evaluate_traj}
	\item \ref{evaluate_energy} and \ref{evaluate_traj} are repeated until the trajectory is finalized (e.g. reached the end time).
\end{enumerate}
Software for conducting dynamics simulations is interfaced to quantum chemistry software.
This interface passes the geometric information to the quantum chemistry software,
runs the gradient and derivative coupling calculations,
and gathers the energy, gradient, and derivative coupling.
These communications are usually done using ASCII text files.
There are many software that are dedicated to nonadiabatic dynamics simulations.
In particular, the program packages {{\sc Newton-X}}\cite{Barbatti2011WIREs,Barbatti2014WIREs} and {{\sc sharc}}\cite{Mai2018WCMS,sharc}
provide an interface to electronic structure software that includes multireference electron correlation methods.
Some of the quantum chemistry softwares have their own semiclassical dynamics codes.
In particular, \textit{ab initio} multiple spawning (AIMS) is implemented in the program package {\sc molpro},\cite{MOLPRO} 
which is used in conjunction with PIC-CASPT2.
Comprehensive reviews on nonadiabatic (semiclassical) dynamics simulation protocols were recently published in \textit{Chemical Reviews}.\cite{CrespoOtero2018CR,Curchod2018CR}

\section{Summary and Future Prospects}\label{sec_conclusion}

In this review, we have discussed the
theory and application of analytical nuclear gradients
for multireference electron correlation methods,
such as MRCI, MRPT, and MRCC.
Since the 1980s, the developments in algorithms and the emergence of highly efficient computer programs
have made it possible to perform many chemical applications using these methodologies.
One of such applications of these gradient algorithms is
structure determination of chemical systems,
which serves as useful tools
in studies of photochemistry and strongly correlated systems.
Moreover, direct nonadiabatic dynamics simulations with multireference electron correlation models have become a practical tool in recent years,
which had previously been deemed unpractical.
Since the mechanism of many photochemical and photophysical processes remains difficult to be investigated experimentally,
the theoretical methods reviewed here are powerful tools for 
studying such processes. 

In spite of the active developments in the past,
there is ample room for improving the computational algorithms,
in particular, for further reducing the computational burden for correlated electronic structure calculations.
Recent progresses in
explicitly correlated algorithms
and their analytical gradients,\cite{Gyorffy2017JCP,Gyorffy2018JCP}
the approximate full CI solvers and their analytical gradients with large active spaces,\cite{Liu2013JCTC,Hu2015JCTC,Nakatani2017JCP}
and the local correlation algorithms\cite{Menezes2016JCP,Guo2016JCP,Kats2019JCP,Frank2017MolPhys,Pinski2019JCP}
have the potential to further expand the role of multireference methods in chemical application.

\section{Biographies}

Jae Woo Park was born in Gwangju, Korea, in 1989. He graduated early from Gyeonggi Science High School in Suwon, Korea in 2007.
He then received his B.Sc. degree in 2010, and Ph.D. degree in 2015, at the Department of Chemistry of
Pohang University of Science and Technology (POSTECH), Pohang, Korea, with Prof. Young Min Rhee [now at Korea Advanced Institute of Science and Technology (KAIST)].
After working as a postdoctoral researcher in the Institute for Basic Science (IBS) Center for Self-assembly and Complexity (CSC) (2015--2016),
he worked as a postdoctoral researcher in Northwestern University, in the laboratory of Prof. Toru Shiozaki (2016--2018).
While at Northwestern, he was involved in developing the CASPT2 derivative coupling code in the program package {{\sc bagel}}.
Since 2018, he has been an assistant professor at Chungbuk National University (CBNU) in Cheongju, Korea.
His research focuses on developing and applying multireference quantum chemistry theories for complex systems.

Rachael Al-Saadon was born in Terre Haute, IN, in 1990.
She received her B.S. in chemistry at Indiana University Bloomington (2012) and Ph.D. in chemistry at Duke University (2018).
Presently, she is a postdoctoral fellow in the group of Prof. Toru Shiozaki at Northwestern University.
She is developing CASPT2 spin--orbit coupling methodology.

Matthew K. MacLeod is from Longmont, CO. He received his B.A. in chemistry from Claremont McKenna College in 2003. His doctorate work under the supervision
of Josef Michl involved exploring singlet excited state potential energy surfaces with stochastic search and gradient based methods in
order to understand exciton localization in oligosilanes. His post-doctoral research (2012--2014) with Prof. Toru Shiozaki at Northwestern on code generation of CASPT2 analytical gradients
provided the theory and tools which aided in the discovery of equilibrium geometries of open shell organic molecules such as the porphyrin radical cation.
Matthew currently builds machine learning models to understand natural language, another multireference problem in life.

Toru Shiozaki is from Fukuoka, Japan. 
Toru earned his PhD in 2010 from the University of Tokyo under the supervision by Prof. So Hirata at University of Florida.
He then worked with Prof. Hans-Joachim Werner as a JSPS fellow at University of Stuttgart (2010--2012), followed by
an assistant professorship at Northwestern University (2012--2019).
Together with Garnet Chan, Toru has recently created a venture-backed deep-tech startup, Quantum Simulation Technologies, Inc. (dba. QSimulate),
aiming to provide advanced quantum mechanical simulation platforms to chemical and pharmaceutical industries. 
Toru is QSimulate's cofounder and full-time CEO starting September 2019.

Bess Vlaisavljevich was born in St. Paul, MN in 1984. She received her B.A. in chemistry from the University of Minnesota Morris in 2007. 
She then received her M.Sc. and Ph.D. degrees working with Prof. Laura Gagliardi from the Department of Chemistry at the University of Minnesota
Twin Cities in 2010 and 2013, respectively. Following this, she worked as a postdoctoral researcher at the University of California Berkeley with 
Prof. Berend Smit and later at Northwestern University with Prof. Toru Shiozaki. While at Northwestern, she was involved in developing the 
MS-CASPT2 analytical gradient code in the program package {{\sc bagel}}. In 2017, she began her independent career at the University of South
Dakota. Her research focuses on using multireference quantum chemistry theories for a variety of problems in transition metal and actinide 
chemistry.

\section{Acknowledgments}
The portion of work done at Northwestern University by all of the contributors was supported by National Science Foundation (ACI-1550481 and CHE-1351598),
Department of Energy, Basic Energy Sciences (DE-FG02-13ER16398), and Air Force Office of Scientific Research (FA9550-18-1-0252).
JWP was supported by the National Research Foundation (NRF) grant funded by the Korean government (MSIT) (Grant 2019R1C1C1003657).
BV was supported by the Department of Energy, Basic Energy Sciences (DE-SC0019463).

\bibliography{chemrev-control,chemrev}
\vfill
\end{document}